%% file: main.tex
\documentclass{egpubl}

\usepackage{eg2026}

\ConferenceSubmission

\usepackage{algorithm}
\usepackage{algpseudocode}
\usepackage{amsmath}
\usepackage{soul}
\usepackage{bm}
\usepackage{xspace}
\usepackage{wrapfig}  %
\usepackage{tikz}
\usepackage{tabularx} %

\definecolor{dgreen}{rgb}{0.0, 0.5, 0.0}
\definecolor{dred}{rgb}{0.65, 0.16, 0.16}

\definecolor{mblue}{rgb}{0.45, 0.82, 0.93}
\definecolor{mgreen}{rgb}{0.55, 0.78, 0.25}
\definecolor{mred}{rgb}{1., 0.39, 0.31}
\definecolor{mpurple}{rgb}{0.74, 0.50, 0.74}

\definecolor{yarnGreen}{RGB}{153, 221, 213}
\definecolor{yarnOrange}{RGB}{221, 167, 103}
\definecolor{yarnRed}{RGB}{199, 108, 108}
\definecolor{diagramOrange}{RGB}{242, 190, 34}

\newcommand{\Description}[1]{} %

\newcommand{\citet}[1]{\cite{#1}}

\newcommand{\Method}{barrier-filtered\xspace}

\usepackage[T1]{fontenc}
\usepackage{dfadobe}

\usepackage{cite}  %
\BibtexOrBiblatex

\electronicVersion
\PrintedOrElectronic

\usepackage{egweblnk} 
\usepackage{lineno}

\title[Unlocking Thickness Modeling for Codimensional Contact Simulation]{Unlocking Thickness Modeling for Codimensional\\Contact Simulation}

\author[G. Gomez-Nogales, Z. Chen, R. Martin, E. Garces \& D. M. Kaufman]
{\parbox{\textwidth}{\centering Gonzalo Gomez-Nogales$^{1,2}$, Zhen Chen$^{3}$, Rosalie Martin$^{2}$, Elena Garces$^{2}$ and Danny M. Kaufman$^{3}$} \\
{\parbox{\textwidth}{\centering $^1$University Rey Juan Carlos, Mostoles, Spain\\$^2$Adobe Research, Paris, France\\$^3$Adobe Research, Seattle, US}}}

\begin{document}

\teaser{
    \centering
    \begin{tikzpicture}
    \node[anchor=south west,inner sep=0] (image) at (0,0){\includegraphics[width=0.97\linewidth]{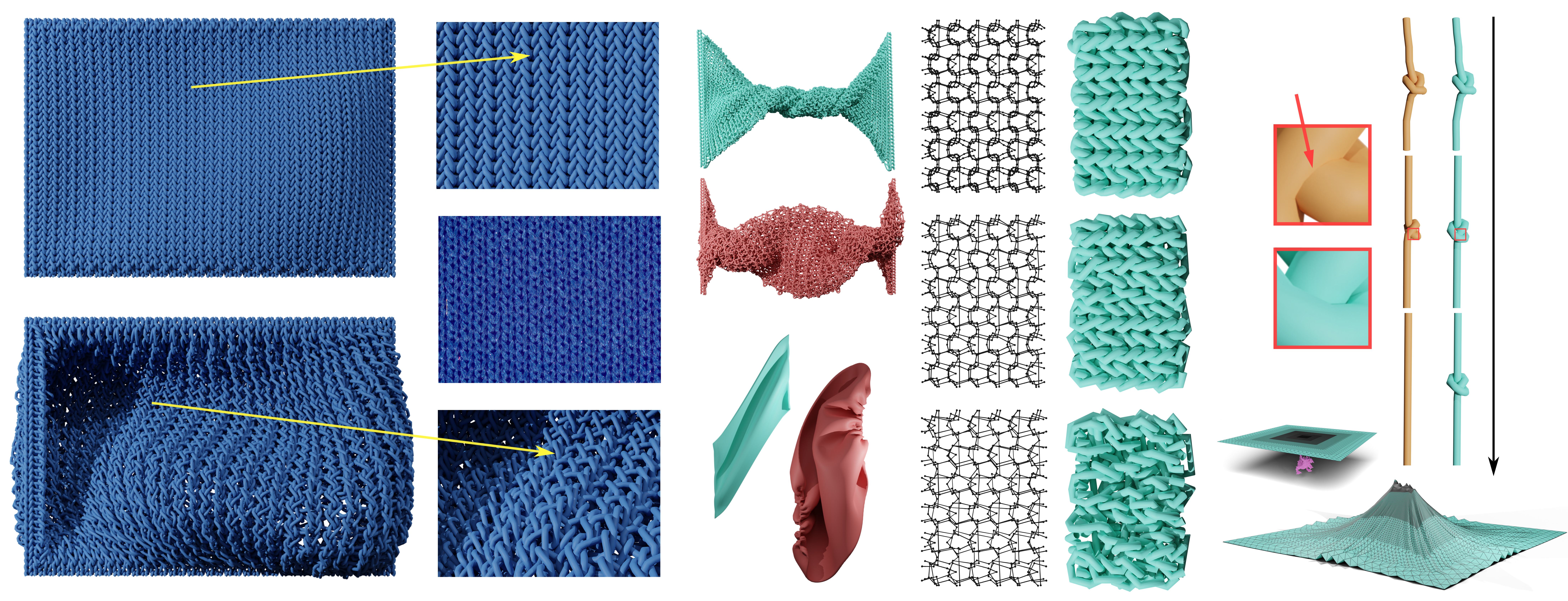}};
    \begin{scope}[x={(image.south east)},y={(image.north west)}]
        \node[anchor=north] at (0.23, 0.02) {{(a)}};
        \node[anchor=north] at (0.5, 0.02) {{(b)}};
        \node[anchor=north] at (0.69, 0.02){{(c)}};
        \node[anchor=north] at (0.9, 0.02){(d)};
        \node[anchor=south] at (0.82, 0.83){\footnotesize{\textcolor{red}{self-intersected}}};
        \node[anchor=south] at (0.875, 0.93){\scriptsize{Culled}};
        \node[anchor=south] at (0.912, 0.93){\scriptsize{Ours}};
        \node[anchor=north] at (0.51, 0.92){\footnotesize{Ours}};
        \node[anchor=north] at (0.51, 0.72){\footnotesize{Barrier}};
        \node[anchor=east] at (0.49, 0.4){\footnotesize{Ours}};
        \node[anchor=east] at (0.51, 0.1){\footnotesize{Barrier}};
        \node[anchor=north] at (0.95, 0.21){\scriptsize{}{Cloth Drape}};
        \node[anchor=south, rotate=90] at (0.01, 0.3) {{Barrier}};
        \node[anchor=south, rotate=90] at (0.01, 0.7) {{Our Method}};
        \node[anchor=south, rotate=-90] at (0.95, 0.32) {\scriptsize{Pulling}};
        \node[anchor=south, rotate=90] at (0.285, 0.5) {\footnotesize{Real Image}};
    \end{scope}
    \end{tikzpicture}
    \caption{Accurate simulation of fabrics and yarns is challenged by real-world material thickness. In (a) we reproduce the DKP yarn pattern from Sperl et al.~\cite{Sperl2022}. Existing barrier methods \textit{(e.g.}, Li et al.~\cite{Li2021CIPC}), ensure \emph{robust}, non-intersecting simulations but introduce severe pattern distortions (a, bottom) due to nonphysical forces between stencils neighbors closer than material thickness. These issues worsen under deformations, such as twists (b, top), and are even more severe in shell-based simulations of thicker fabrics (b, bottom). A common workaround to cull troubling contacts~\cite{Kaldor2008}, produces unacceptable self-intersections that break pattern designs (d, top). We address these challenges with a practical Barrier-Filtering method that enables all resolutions (c), supporting controllable, thickness-independent accuracy. Our approach achieves artifact-free results even with irregular or strongly graded meshes (d, bottom).}
    \label{fig:teaser}
}

\maketitle
\begin{abstract} 
   In this work we analyze and address a fundamental restriction that blocks the reliable application of codimensional yarn-level and shell models with thickness, to simulate real-world woven and knit fabrics. 
   As discretizations refine toward practical and accurate physical modeling, such models can generate non-physical contact forces with stencil-neighboring elements in the simulation mesh, leading to severe locking artifacts. While not well-documented in the literature, this restriction has so far been addressed with two alternatives with undesirable tradeoffs. One option is to restrict the mesh to coarse resolutions, however, this eliminates the possibility of accurate (and consistent) resolution simulations across real-world material variations. A second alternative instead seeks to cull contact pairs that can create such locking forces in the first place. This relaxes resolution restrictions but compromise robustness. Culling can and will generate unacceptable and unpredictable geometric intersections and tunneling that destroys weaving and knitting structures and cause unrecoverable pull-throughs. 
   We address these challenges to simulating real-world materials with a new and practical contact-processing model for thickened codimensional simulation, that removes resolution restrictions, while guaranteeing contact-locking-free, non-intersecting simulations. 
    We demonstrate the application of our model across a wide range of previously unavailable simulation scenarios, with real-world material yarn and fabric parameters and patterns, challenging simulation conditions and mesh resolutions, and both rod and shell models, integrated with the IPC barrier.


\end{abstract}

\input{01-introduction}

\input{02-related-work}

\input{03-method}

\input{04-evaluation}

\input{05-conclusion}

\bibliographystyle{eg-alpha-doi} 
\bibliography{bibliography}
\clearpage
\appendix 
\input{0A-implementation-details_appendix}

\input{0B-more-comparisons_appendix}

\end{document}

%% file: 01-introduction.tex
\section{Introduction}
\label{sec:intro}
Codimensional models are popular, accurate, and efficient representations for simulating the physical behavior of thin materials. The \emph{material} properties of such rod and shell structures are captured by incorporating their thicknesses (codimensional offsets) into the computation of each model's constitutive behavior (\textit{i.e.}, moduli) and mass. Correspondingly, the complex \emph{geometric} interactions \emph{between} codimensional domains, as in knits, yarn-level materials and self-contacting fabrics, are modeled by contact-processing with the offset surfaces defined by these same codimensional thicknesses. This framework enables efficient simulations that can predictably capture the constitutive and contact behaviors of complex structures comprised of thin volumes, via lower degree-of-freedom (DOF) computations with discretized curves and surfaces. 

In this work, we address a fundamental restriction that limits the reliable use of these codimensional models for accurately simulating real-world fabrics. %
Refining discretizations for higher accuracy inevitably leads these models to introduce non-physical contact forces between stencil-neighboring mesh elements, as we show in Figures~\ref{fig:diagram_barrier_contact_locking} and \ref{fig:show_pushing_artifacts_toy_example}.
These interactions produce severe non-physical contact-locking artifacts created by contact forces expanding neighboring elements' materials outwards to artificially satisfy distance bounds imposed by the material thickness. In turn, this is resisted by elastic forces generating inflated and often artificially stiffened (even close-to-rigid) simulation of soft materials. The situation is only exacerbated when strain-limiting forces~\cite{English2008, Chen2010DevelopableCloth, Wang2010, Jin2017, Wang:2021:GBS, Li2021CIPC}, standard in codimensional simulations, are employed as well; see e.g., Figure\ \ref{fig:cloth_on_sphere}. 
While not well documented in the literature, this restriction has so far been addressed with two alternatives with undesirable tradeoffs.

\setlength{\columnsep}{0.5em}
\setlength{\intextsep}{0em}
\begin{wrapfigure}{r}{70pt}
   \includegraphics[width=70pt]{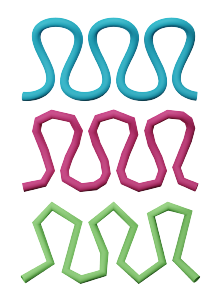}
   \label{fig:discretization-example-inset}
\end{wrapfigure}

As a first option, many methods simply restrict themselves to coarser meshes that avoid this contact-based locking \cite{Li2021CIPC,Ando2024}. However, this eliminates the possibility of accurate (and consistent) resolution simulations across material variations; see inset and Figure\ \ref{fig:resolution_decimation_example}.
We note that this contact locking is in contrast to \emph{membrane locking} artifacts that are mitigated by strain-limiting methods\ \cite{Li2021CIPC,Ando2024} and \emph{increasing} resolution. Contact locking, on the other hand, is reduced by coarsening meshes and so \emph{decreasing} resolution.

A second alternative instead seeks to cull contact pairs that can create such locking forces in the first place \cite{Kaldor2008, Kaldor2010, Leaf2018, Sperl2022}. This latter approach eliminates some mesh-resolution restrictions at the cost of reliability and robustness. Culling can and will generate unacceptable and unpredictable geometric intersections and tunneling that break knit structures and cause unrecoverable pull-throughs. See \textit{e.g.}, Figure~\ref{fig:knots_pull_through_example}. %
\input{images/tex/contact_locking_diagram}
\input{images/tex/self_pushing_toy_example.tex}

We address these challenges to simulating real-world materials with a new contact-processing model for thickened codomain simulation that both removes resolution restrictions and guarantees locking-free, non-intersecting simulations. To do so, we propose an exceedingly simple, safe-filtering approach to efficiently detect and resolve all potential collision-stencil interactions, without contact locking and without intersection. 

In a nutshell, we extend Kaldor et al.'s \shortcite{Kaldor2008} original culling strategy by first carefully identifying and labeling, per simulation mesh, \emph{just} the pairwise contact stencils that are closer than applied thickness in parametric space (both in 1D for rods and 2D for shells). Importantly, this allows us to carefully and conservatively filter both nonuniform discretizations (as applied for rods and shells) and unstructured meshes (as commonly employed in shell-simulation applications). Then, rather than culling out and eliminating these stencils from collision detection and contact processing, we reduce these pairs' effective thickness terms (solely as applied in contact-processing) so that artificial forces are not generated between nearby stencils (\textit{e.g.}, at rest). This carefully leaves the \emph{constitutive} thickness behavior of the codimensional model unchanged and eliminates contact-based locking artifacts from thickness modeling. In turn, this allows simulations to utilize mesh resolutions with sufficient resolution to capture curvature in rods and shells, while preserving \emph{geometric} thickness modeling almost everywhere (with consistent frictional contact behavior), and ensuring non-intersection for all simulation scenarios, irrespective of collision speeds or contact-configurations. 

In Section \ref{sec:method}, we detail the preprocessed construction and runtime application of our underlying filtering model, detail its application to both rod and shell models, and then cover the necessary steps for its efficient integration into an IPC-based \cite{Li2020IPC} hybrid simulator for both yarn-level simulation and fabrics. In Section\ \ref{sec:evaluation}, we then carefully analyze, compare, and demonstrate our filtering models' ability to accurately and robustly capture intersection-free simulations, independent of resolution, across an unrestricted range of real-world-reported yarns and fabric material thicknesses, and yarn-pattern configurations.

\vspace{1em}

\input{images/tex/resolution_decimation_example}

\input{images/tex/knots}

%% file: images/tex/contact_locking_diagram.tex
\begin{figure}[tb]
  \centering
  \begin{tikzpicture}
    \node[anchor=south west,inner sep=0] (image) at (0,0){\includegraphics[width=0.95\linewidth]{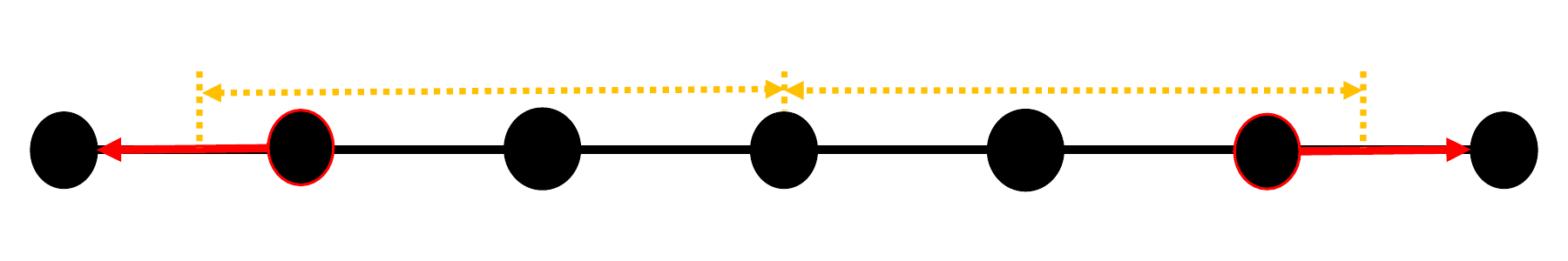}};
    \begin{scope}[x={(image.south east)},y={(image.north west)}]
        \node[anchor=north] at (0.5, 0.3) {$V_0$};
        \node[anchor=north] at (0.655, 0.3) {$V_1$};
        \node[anchor=north] at (0.81, 0.3) {$V_2$};
        \node[anchor=north] at (0.345, 0.3) {$V_3$};
        \node[anchor=north] at (0.19, 0.3) {$V_4$};
        \node[anchor=north] at (0.1, 0.4) {\textcolor{red}{$F_C$}};
        \node[anchor=north] at (0.9, 0.4) {\textcolor{red}{$F_C$}};
        \node[anchor=south] at (0.3, 0.6) {\textcolor{diagramOrange}{$a$}};
        \node[anchor=south] at (0.7, 0.6) {\textcolor{diagramOrange}{$a$}};
    \end{scope}
    \end{tikzpicture}
  \caption{\textbf{Contact-based locking} occurs for standard codimensional barrier models when local resolution in the reference mesh (a simple polyline midline segment in this example), has distances for contact stencils, here e.g., between vertex $V_0$, and vertices $V_2$ and $V_4$, below the activation threshold, $a$. In turn, this generates nonphysical expansive contact forces, $F_C$ for these stencils, leading to inflated domains (even at rest),  artificially stiff material responses, and nonzero rest forces. See Figure\ \ref{fig:show_pushing_artifacts_toy_example}.}
  \label{fig:diagram_barrier_contact_locking}
\end{figure}

%% file: images/tex/self_pushing_toy_example.tex
\begin{figure}[tb]
  \centering  
  \includegraphics[width=\linewidth]{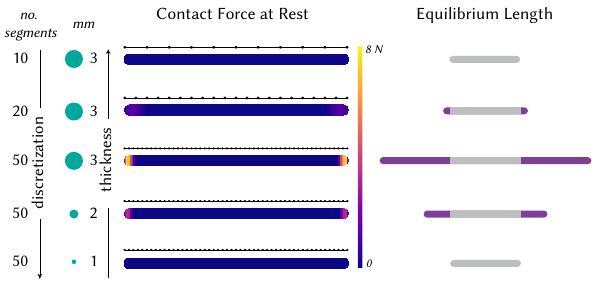}  
  \caption{\textbf{Contact locking in codimensional simulations.} Gravityless equilibrium of a 5cm yarn with only contact forces enabled. %
  Left: contact forces at rest when varying resolution (top, fixed thickness) or thickness (bottom, fixed resolution). Note that, at rest, no contact forces should be active. Right: due to nonphysical forces (see Figure \ref{fig:diagram_barrier_contact_locking}), the yarn expands beyond its rest length when the resolution is too fine for the chosen thickness. This limits both feasible resolution and achievable accuracy.
  }
  \label{fig:show_pushing_artifacts_toy_example}
\end{figure}

%% file: images/tex/resolution_decimation_example.tex
\begin{figure}[ht]
    \includegraphics[width=\linewidth]{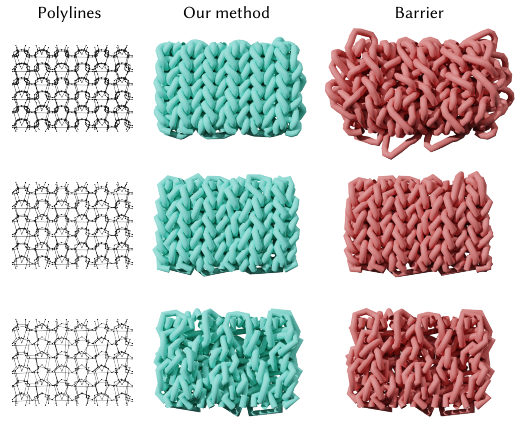}
    \caption{\textbf{Resolution vs. material thickness.} In barrier models, yarn thickness limits the maximum resolution achievable without artifacts. Here, we simulate the DKP pattern with a 150D/72F Polyester yarn\ \cite{Sperl2022} using our filtered barrier (green) and the standard barrier\ \cite{Li2021CIPC} (red), from coarse to fine resolution (bottom to top; discretizations at left). Bottom: at coarse resolution both methods agree, but accuracy is low.  Middle: the barrier method exhibits locking artifacts (expansions and gaps), while our method remains stable. However,  the resolution is too coarse for many applications. Top: at finer resolution, our filtered barrier models the real pattern smoothly, whereas locking completely destroys the standard barrier solution.}
    \Description{Resolution Decimation Example}
    \label{fig:resolution_decimation_example}
\end{figure}

%% file: images/tex/knots.tex
\begin{figure}[h]
    \begin{tikzpicture}
    \node[anchor=south west,inner sep=0] (image) at (0,0){\includegraphics[width=\linewidth]{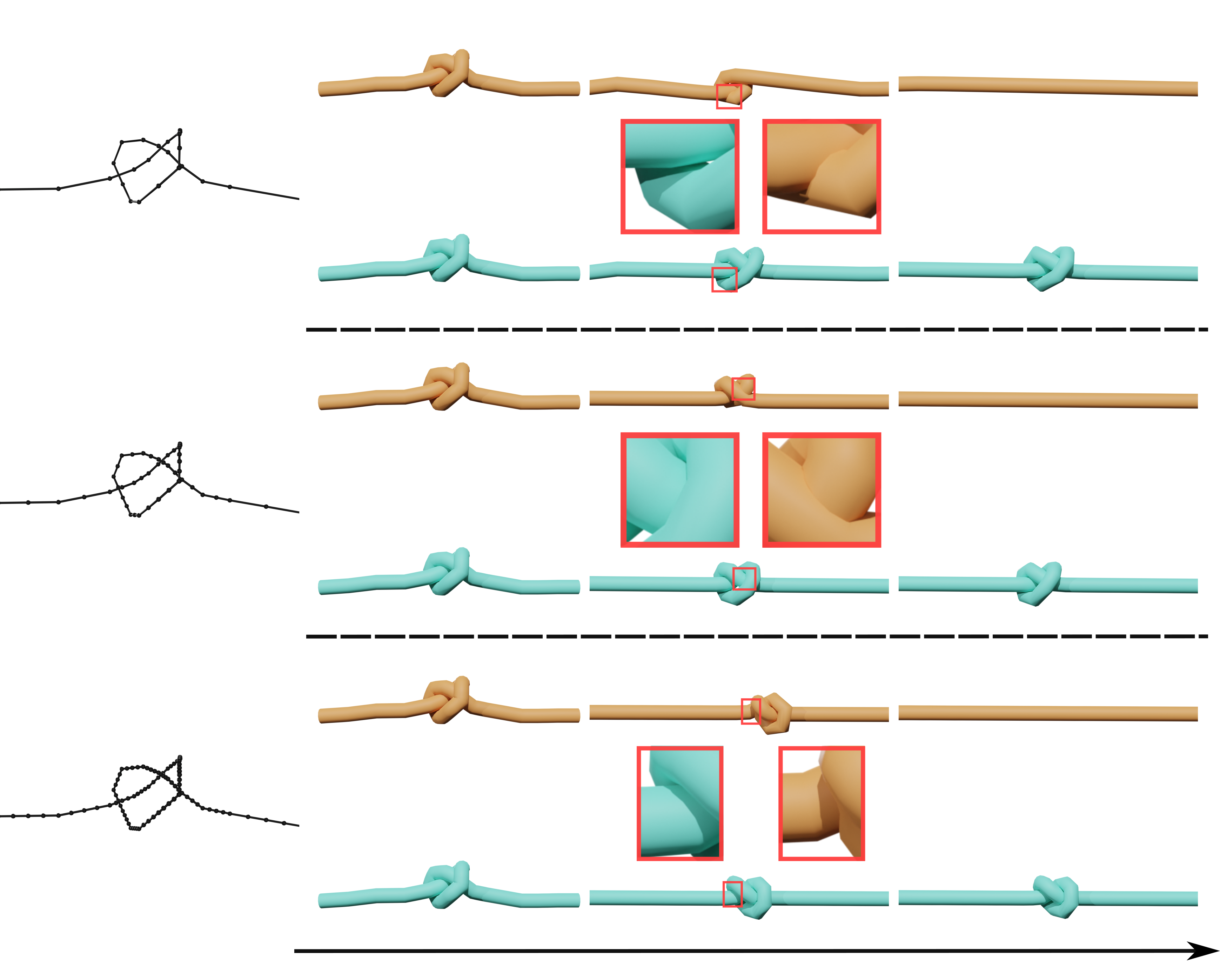}};
    \begin{scope}[x={(image.south east)},y={(image.north west)}]

        \node[anchor=south] at (0.12, 0.1) {\footnotesize{120 segments}};
        \node[anchor=south] at (0.12, 0.42) {\footnotesize{60 segments}};
        \node[anchor=south] at (0.12, 0.74) {\footnotesize{30 segments}};
        \node[anchor=north] at (0.6, 0.03) {\footnotesize{Simulating}};

        \node[anchor=north] at (0.22, 0.115) {\footnotesize{Our}};
        \node[anchor=north] at (0.22, 0.30) {\footnotesize{Culled}};
        \node[anchor=north] at (0.22, 0.435) {\footnotesize{Our}};
        \node[anchor=north] at (0.22, 0.62) {\footnotesize{Culled}};
        \node[anchor=north] at (0.22, 0.755) {\footnotesize{Our}};
        \node[anchor=north] at (0.22, 0.94) {\footnotesize{Culled}};
    \end{scope}
    \end{tikzpicture}
\caption{\textbf{Knot test.} A 0.3 mm thick knotted yarn is pulled from both ends. We compare our barrier-filtered simulation (\textcolor{yarnGreen}{\textbf{green}}) with a contact-culled simulation (\textcolor{yarnOrange}{\textbf{orange}}) ~\cite{Kaldor2008}, using the smallest culling radius that removes expansive contact forces. 
Left: initial polyline knot configurations at different resolutions. As the knot is pulled, all culled simulations fail—the knot breaks apart due to intersections and pull-throughs, regardless of resolution or culling radius. In contrast, our barrier-filtering preserves the knot’s tightened structure at all resolutions, free of expansion artifacts.
}
  \Description{Examples with knots to show pull-through under same culling window vs our filtered barried method}
  \label{fig:knots_pull_through_example}
\end{figure}

%% file: 02-related-work.tex
\section{Related Work}
\label{sec:related-work}

\paragraph*{Thin Shell Cloth Simulation}
Since the pioneering work of Terzopoulos and Fleischer~\cite{Terzopoulos1988}, cloth simulation has been a longstanding focus in both computational mechanics and computer graphics. Numerous advances have been made, including the adoption of implicit time integration~\cite{Baraff1998}, adaptive remeshing techniques~\cite{Grinspun2002, Narain2012ARCSim, Koh2014VDA}, multiresolution methods~\cite{Zhang2022, Zhang2023, Zhang2024progressive, Chen2021, Chen2023CWF, Wang2018, Xian2019}, GPU acceleration~\cite{Tang2013, Li2020Pcloth, Wang:2021:GBS, Wu:2022:GBM, Huang2024:GIPC, Ando2024}, and robust contact and collision handling strategies~\cite{Tang2016, Tang2018, Li2020IPC, Li2021CIPC, Ando2024, Huang2024:GIPC}.

\paragraph*{Yarn-level Cloth Simulation}
To capture yarn-level dynamics, each yarn thread is typically modeled as an elastic rod~\cite{Bergou2008, Bergou2010, Dimitri2007}, with yarn-yarn contact driving fabric deformation. This line of research was initiated by Kaldor et al.~\cite{Kaldor2008}, who used cubic B-splines to model individual yarns. Kaldor et al.\cite{Kaldor2010} later introduced optimizations for efficiently updating nonlinear contact terms. The persistent contact model of Cirio et al.~\cite{Cirio2014, Cirio2015} improved simulation efficiency by discretizing interlaced yarns through yarn crossings and sliding. Sánchez-Banderas et al.~\cite{Banderas2020} extended this idea to handle both intra- and inter-layer contacts implicitly.

Casafranca et al.~\cite{Casafranca2020} proposed a hybrid cloth model combining yarns and triangles to balance the efficiency of triangle-based simulation with the rich nonlinear and plastic effects of yarn models. A GPU-based yarn simulator was introduced in Leaf et al.~\cite{Leaf2018} to interactively design periodic yarn patterns. Sperl et al.~\cite{Sperl2020Homogenized, Sperl2021} explored the homogenization of yarn-based models into thin shells and further developed a method to estimate yarn-level parameters from real fabric data~\cite{Sperl2022}.

Most recently, Zhang et al.~\cite{Zhang2024} presented a method for determining yarn model parameters from experimentally measured Young's moduli via an extended homogenization framework, aligning yarn-level and shell-level hyperelastic energies under various surface deformations. Yuan et al.~\cite{Yuan2024} introduced an approach to homogenize yarn-level dynamics into a volumetric enclosure with a volume-preserving constraint.

\paragraph*{Modeling Material Thickness}

In the above-covered works, the material properties of codimensional structures directly incorporate thicknesses in the computation of each model's constitutive behavior, \textit{e.g.}, bending behavior, and mass. Alternatively, solid shell formulations\ \cite{Hauptmann1998} augment surface models with  translational degrees of freedom across the thickness dimension. This is a widely applied and flexible approach\ \cite{HARNAU2002805, Chen2023:thick-shell, Montes2023}, but fundamentally, such volumetric discretizations, such as linear prism elements, often suffer from severe locking—particularly under strong compression and plastic deformation~\cite{Doll2000} 
Higher-order elements~\cite{Montes2024q3t} can mitigate, although not eliminate, these issues, at the expense of increased complexity and computational cost.

\paragraph*{Modeling Geometric Thickness}
In turn, to model the geometric interactions between thickened contacting codimensional models (both for self and external contacts), knitted yarn and fabric methods have focused on applying barrier-based contact models. Alternately, constraint-based contact methods have also applied similar strategies by offsetting constraints\ \cite{Narain2012ARCSim, Li2018:ARGUS}. However, they are applied to improve constraint feasibility, and do not provide consistent thickness modeling nor stable contact behavior\ \cite{Li2021CIPC}. 

For barrier-based models, with an initial focus on yarns, geometric thickness for codimensional models has typically been modeled by assigning a thickness offset to the codimensional model, \textit{e.g.}, an offset radius to each yarn centerline. To enforce offset thicknesses between yarns, Kaldor et al.~\cite{Kaldor2008} propose a yarn-yarn barrier-based collision model based on B-spline centerlines, evaluating interactions between sampled curve segments with a simple culling strategy that ignores nearby collisions between contact stencils. This model was later accelerated using BVH-based pair filtering~\cite{Kaldor2010}, and extended to GPU for interactive pattern design~\cite{Leaf2018}. Cirio et al.~\cite{Cirio2014, Cirio2017} extended the method in Sueda et al.~\cite{Sueda2011} for yarn fabrics based on explicit knowledge of yarn patterns. Sperl et al.~\cite{Sperl2022} enhanced the robustness of Kaldor's formulation by replacing its original barrier potential with the IPC-based barrier function introduced by Li et al.~\cite{Li2020IPC}. Li et al.~\cite{Li2021CIPC} introduced a barrier-based thickness model for both shells and elastic rods that extends the IPC contact-barrier to capture a biphasic thickness modeling both a softer outer-range response, and an innner-range dense core, that disallows intersection entirely. Across all such barrier based methods, a fundamental limitation is the contact-locking behavior we cover below and in the following sections: barriers introduce unavoidable and undesirable self-pushing artifacts when mesh resolution is finer than modeled thickness (see Figure\ \ref{fig:show_pushing_artifacts_toy_example}), that can not be safely eliminated by culling (see \textit{e.g.}, Figures  \ref{fig:knots_pull_through_example}, and \ref{fig:twisting_A1_small_comp_DF_CIPC_KALDOR_11}).

\paragraph*{Membrane Locking and Contact Locking.}
Membrane locking refers to the phenomenon in which thin shell models exhibit artificially stiff responses under bending, particularly when using low-order finite elements. This issue has been thoroughly investigated by Quaglino~\cite{quaglino2012membrane, Quaglino2016}, who introduced several benchmark tests to characterize locking behavior in triangle-mesh-based kinematic models. Various approaches have been proposed to address membrane locking, including nonconforming elements~\cite{English2008}, adaptive mesh refinement~\cite{Narain2012ARCSim}, isometric constraint enforcement~\cite{chen2019locking}, and strain-limiting methods~\cite{Li2021CIPC}. In general, membrane locking can be mitigated by increasing mesh resolution.

In contrast, contact locking, to the best of our knowledge, has received limited attention in the literature. Contact locking arises when the element size in a finite element system (\textit{e.g.}, triangle meshes for thin shells or elastic rods for yarns) becomes \emph{smaller} than the contact threshold. In such cases, contact forces affect neighborhoods beyond the element itself (Figure\ \ref{fig:diagram_barrier_contact_locking}), producing undesirable self-pushing behavior and resulting in an artificially stiff material response. This effect is further amplified when strain-limiting methods~\cite{Li2021CIPC,Ando2024} are applied, as they restrict deformations that would otherwise relieve contact-induced compression. As a result, the material appears even stiffer. Unlike membrane locking, contact locking can be alleviated by \emph{decreasing} the resolution—an approach that, however, exacerbates membrane locking and inaccurate simulation, highlighting a fundamental and undesirable trade-off.

%% file: 03-method.tex
\section{Filtered Barriers for Simulating Thickened Codomains}
\label{sec:method}

As covered above in Section\ \ref{sec:related-work}, codimensional simulation solutions for modeling geometric thickness in yarns and fabrics have focused on applying barrier-based contact models. While constraint-based contact methods have applied similar concepts of thickened constraint ``offsets''\ \cite{Narain2012ARCSim}, they are purposed for improving contact enforcement and do not provide consistent thickness modeling\ \cite{Li2021CIPC}. 

Specifically, thickness for codimensional modeling has been incorporated via contact barrier energies, $b(d,a)$, evaluated at distances, $d$, parameterized by activation distances, $a$. These activation distances are where a barrier's repulsion forces initially become non-zero, and so are set to match the material thickness's offset from a rod's (respectively shell's) midline (respectively midsurface). Here and in the following, we use $h$ to denote material thickness. Assuming (as standard in prior yarn work) cylindrically thickened rods, offsets are then $h/2$ for both yarns and shells. Choice of barrier function then enables different effective constitutive contact-compressive behaviors for yarns and fabrics\ \cite{Kaldor2008, Sperl2022, Li2021CIPC}. 

In turn, thickened codomain simulation methods model with both spline-curve \cite{Kaldor2008, Kaldor2010, Leaf2018} and polyline\ \cite{Sperl2022, Li2021CIPC} discretization's for yarn midlines, and generally apply triangulated meshes for shell midsurfaces\ \cite{Li2021CIPC}. Irrespective of base discretization and barrier energy, across methods barriers are correspondingly evaluated on distances between discrete, piecewise stencil pairings: point (quadrature samples in the case of splines) or point/edge pairings for rods, and point/edge/triangle pairings for shells. In this setting, the discrete activation distance for barriers applied for same-material (\textit{e.g.}, self-contacting) pair stencils is set to $h$.

\subsection{Contact-Based Locking}

With the above framework in place, contact-based locking from geometrically modeling thickening occurs whenever (due to local resolution of a discretization) the \emph{reference} distance, $\bar{d}_{i,j}$, for a contact stencil element formed between nearby mesh primitives $i$ and $j$, in the reference (i.e., undeformed) mesh,
is below the activation threshold, $a = h$, defined by the material thickness. See Figure~\ref{fig:diagram_barrier_contact_locking}. When activated, the stencil's contact force causes nonphysical expansion artifacts in the mesh (as discussed in Section~\ref{sec:related-work}), leading, in turn, to an artificially stiff response that resists bending. As a consequence, locking artifacts emerge in the simulation. Note that we differentiate these cases from configuration-dependent intersections where geometric regions are initialized (by rigid transforms and/or deformations) into an intersection and so generate repulsive forces at the start of the simulation. We focus on the resolution of the former unavoidable locking issue here, and refer to Section\ \ref{sec:resolve-intersection} below for more discussion and treatment of this latter, orthogonal, scene-initialization-dependent intersection issue. 
Our filtered collision processing to address this locking then follows from our above observations in a simple and natural fashion, which we next cover below. We then, in the following sections, cover technical details for its efficient application in the C-IPC model, with non-intersection guarantees, for thickened rod and shell simulation with the IPC barrier. 

\subsection{Filtered Collision Processing}

In a precomputation phase we first traverse mesh primitive pairings, $\{i,j\}$, in the set of possible contact stencils, $\mathcal{K}$, \emph{within} each connected mesh component, evaluating their reference distances, $\bar{d}_{i,j}$. If all distances are above the activation threshold, $\bar{d}_{i,j} \geq a(=h)$, $\forall \{i,j\} \in \mathcal{K}$, there is no possible contact locking and we are done. Otherwise, we simply apply the smallest stencil distance, $\bar{d}_{min} = \min_{\{i,j\} \in \mathcal{K}} \bar{d}_{i,j}$, as a new, corrected barrier activation distance, applied just for contact stencils in the reference mesh below the full thickness's assumed activation threshold. That is, given a mesh with deformed positions, $x$, we now apply during simulation a safely filtered barrier 
\begin{align}
\label{eq:single-filter-barrier}
\tilde{b}(d_{i,j}(x), h) = 
\begin{cases}
            b\big(d_{i,j}(x), h\big), \> \bar{d}_{i,j} \geq h,\\
            b\big(d_{i,j}(x), \bar{d}_{min}\big),  \> 0 < \bar{d}_{i,j} < h,
\end{cases}
\end{align}
where $d_{i,j}(x)$ evaluates the distance between deformed mesh primitives $i$ and $j$ at $x$. Although different stencil pairs use different activation thresholds, each pair’s barrier function remains fixed and consistent throughout the simulation.  Application of our filtered barrier, $\tilde{b}(\cdot)$, ensures continued application of the material's thickness \emph{for all} contact interactions, except for \emph{self-contact} interactions within small regional neighborhoods that would otherwise cause the above-covered locking artifacts. Then, solely for local self-contact interactions, within these small pre-identified patches, the geometric thickness imposed by the barriers is safely reduced to ensure that barrier forces continue to repulse the codimensional geometries away from intersection, while avoiding contact locking, and applying a consistent, conservative thickness of $\bar{d}_{min}$.

\subsection{Application to IPC}

Following recent prior work \cite{Sperl2022, Li2021CIPC}, we implement our filtered collision processing for yarns and fabrics with the IPC barrier\ \cite{Li2020IPC},
\begin{align}
b(d, a)=
\begin{cases}
-(d-a)^2 \ln ( d/a ), \> 0<d<a, \\ 
0, \> d \geq a.
\end{cases}
\end{align}
In order to better capture the biphasic behavior of thin materials under contact compression both Sperl et al.~\cite{Sperl2022} and Li et al.~\cite{Li2021CIPC} correspondingly extend the IPC barrier into two modes. A softer response is applied for contacts early into the activation range, corresponding to loose fiber interactions, and a stiffer response is invoked as contact stencils are evaluated with distances deeper in the range, corresponding to fibers in a braid or weave collapsing into dense arrangements with little slack. 

When capturing this biphasic behavior we follow Li et al.'s \shortcite{Li2021CIPC} biphasic model that decomposes the activation range as $a =  \eta + \hat{d}$, modeling a softer outer repulsive response for contact stencil distances, $d$, in the range of $a-\eta < d < a$, and a fully compressed core thickness beyond this, that disallows contact distances drawing any closer. To do so, the above IPC barrier is generalized\ \cite{Li2021CIPC} as,  
\begin{align}
b(d, a, \eta) = b(d - \eta, a - \eta),
\end{align}
so that the IPC barrier now diverges when contact stencils reach a distance of $\eta$ between midline or midsurface primitives. Likewise, note that we retrieve the single-phase barrier behavior with $\eta =0$. 

For cases with $\eta > 0$, we need extra steps in our filtered collision processing. If $\bar{d}_{min} > \eta$ we could possibly proceed as above. However, this is often not satisfied. Moreover, even when this condition is satisfied, safe-filtering could still generate very small effective barrier activation ranges of $\hat{d} = \bar{d}_{min} -\eta$, leading to poorly conditioned and unnecessarily challenging numerical solves\ \cite{Li2020IPC}. At the same time, when it comes to continuous collision detection (CCD) queries, critical for ensuring IPC's non-intersection guarantees, this approach is also problematic since CCD must conservatively bound displacements with respect to the offsets, $\eta$. Already numerically challenging, naively doing so after safe-filtering could lead to degenerate CCD queries, or even failures if the offset, $\eta$, applied in CCD, is not corrected.

To resolve these issues, for $\eta > 0$, when $\bar{d}_{min} < h$, we reduce the filtered barrier to a single-phase mode (with an effective application of $\eta=0$) for just contact stencils $\{i,j\} \in \mathcal{K}$ with $\bar{d}_{i,j} < h$. The motivation being that 1) we expect tight local bending if these stencils are activated, and so greater compression for self-contacts within these small filtered regions, with a necessarily smaller effective thickness, and 2) this correspondingly enables numerically stable solves with the filtered barrier without concern for the size of $\hat{d}$ and $\eta$. 

In summary, for single-phase thickness simulation, with $\eta = 0$, we use the above-covered, single-filtered barrier in Equation (\ref{eq:single-filter-barrier}). Then, for biphasic thickness simulation, with $\eta > 0$, we apply the double-filtered barrier
\begin{align}
\label{eq:double-filter-barrier}
\tilde{b}(d_{i,j}(x), h, \eta) = 
\begin{cases}
            b\big(d_{i,j}(x) , h,  \eta \big), \> \bar{d}_{i,j}  \geq h,\\
            b\big(d_{i,j}(x), \bar{d}_{min}, 0 \big),  \> 0 < \bar{d}_{i,j} < h.
\end{cases}
\end{align}

Finally, for both one- and two-phase barrier applications, we build and query spatial-hash acceleration structures for collision detection using the full, unmodified activation distances (and offsets as appropriate). While this generates overly conservative broad-phase queries for filtered collision stencils (unnecessary false positives), this enables the direct use of standard, off-the-shelf, well-optimized spatial hashing codes (likewise BVH if desired instead). In future work, it could then be interesting to consider further acceleration via customized, per-stencil hashing, but we do not  anticipate significant speedups. We then apply the robust ACCD method \cite{Li2021CIPC} for all CCD queries with appropriately updated offsets, as applicable.

\subsection{Resolving Input Configuration Intersections}
\label{sec:resolve-intersection}

As covered above, our barrier-filtering addresses contact-locking, which is generated by contact-stencil forces expanding neighboring elements’ outwards to artificially satisfy distance bounds imposed by  material thickness. This is in contrast to when, either by editing or input design-pattern errors, non-neighboring geometric regions of yarns or fabrics (including from other yarns or fabric components) are initialized closer than the material thickness and so into intersection. 

We note, however, that barrier-filtering naturally resolves such initially infeasible configurations by construction. Here, as we build filtered contact-stencil pairs by analyzing distances along polylines and triangulations, these latter type of pairs, brought into intersection by design pattern arrangement \emph{are not filtered}. To resolve these intersections we can simply initialize any such pattern with our barrier-filtering, see examples in Section\ \ref{sec:evaluation} below, by first simulating the pattern to equilibrium prior to any additional modeling. In doing so, barrier forces naturally push intersections outwards so that remaining barrier compressions are then in balance with the materials' elastic forces and any imposed boundary conditions. 

We caution, however, that there are two cases of pattern-design intersection failures that this approach \emph{will not and does not} address. The first is \emph{entanglement}: if meshed centerlines or midsurfaces intersect in an input pattern, applied barrier forces will arbitrarily push the pattern apart and so can not be expected to correctly detangle to a reasonable design. The second is core-material intersection: if Li et al.'s\ \cite{Li2021CIPC} hard thickness offset, $\eta$, (covered above) is applied, and starting configurations intersect within this offset (with contact-stencil primitives brought closer than $\eta$), barrier-filtering will likewise not be able to detangle and push apart the input to a reasonable starting configuration for the design. Here a reasonable option, via an easy extension of our current barrier-filtered approach, would be to apply an additional, temporary filter for all such additionally detected non-neighboring contact-stencil pairs, temporally setting their hard-barrier thickness offset, $\eta$, to 0, until after relaxation to equilibrium pushes them far enough apart to satisfy their targeted $d > \eta$ constraint.

%% file: 04-evaluation.tex
\section{Evaluation}
\label{sec:evaluation}
We evaluate our contact-processing model by analyzing its ability to 1) eliminate contact-locking artifacts, 2) avoid the tunneling issues of culling methods, and 3) maintain accurate, intersection-free simulations across a diverse range of fabric materials and discretizations. We evaluate our method and compare with previous alternatives with both yarn-level and shell-based simulations. Implementation details, parameters used, and additional evaluations are covered Appendix~\ref{sec:appendix-implementation} and~\ref{sec:appendix-additional-comparisons}.%

\subsection{Yarn Pattern Evaluations}

We first evaluate yarn-level patterns in two settings: static equilibrium solves and dynamic deformation exercises.

\paragraph*{Yarn-Pattern Equilibria}
When initializing yarn-level patterns, the pattern is first \emph{relaxed} by simulation to its equilibrium state. With standard barrier methods for resolving thickness the pattern can undergo large, unstable deformations, resulting in significant distortions. In Figure~\ref{fig:real_world_yarns_comp}, we compare our method with the barrier method~\cite{Li2021CIPC} against real-world captures of the same patterns. Here, we render the simulated results using the corresponding real-world color style. Our method closely matches the physical equilibria patterns, while the barrier method again produces noticeable discrepancies. It is worth noting here, that while the culling method proposed by Kaldor et al.~\cite{Kaldor2008} can mitigate these artifacts to some extent, it can not eliminate them. As shown in Figure~\ref{fig:toy_example_kaldor_self_pushing_forces}, culled simulations can still produce undesirable expansion artifacts even with aggressive culling radii. Additional comparisons can be found in Figure~\ref{fig:yarn-level-equilibria} in Appendix~\ref{sec:appendix-additional-comparisons}, where our method successfully brings the input patterns to equilibrium without distortion, while barrier methods generates significant visible artifacts.

\input{images/tex/real_world_yarn_comp_closeup.tex}
\input{images/tex/self_pushing_with_Kaldor}

\paragraph*{Exercising Yarn Patterns}
In Figure~\ref{fig:twisting_A1_small_comp_DF_CIPC_KALDOR_11}, we compare our method in dynamic simulations with a direct barrier formulation~\cite{Li2021CIPC} and the culling approach of Kaldor et al.~\cite{Kaldor2008} on the \emph{A1} pattern from Sperl et al.~\cite{Sperl2022}. As shown, the barrier method exhibits self-expansion artifacts early on, which are effectively resolved by our approach. Meanwhile, although the culling method also avoids self-pushing, it tends to over-cull self-contacting yarn pairs, resulting in visible intersections in the zoomed-in views. In Appendix~\ref{sec:appendix-additional-comparisons} (Figure~\ref{fig:our_barrier_comparison}), we further demonstrate results on additional motions, including stretching and shearing of the same patch as in~\cite{Li2021CIPC}. Once again, our method achieves material-consistent deformations, whereas the barrier-based approach suffers from self-expansion artifacts caused by contact locking. These artifacts manifest as noticeable separation, gapping in the pattern, and unnatural distortions in the final shapes.

\input{images/tex/twist_comp_yarn.tex}

\subsection{Shell Fabrics Evaluation}
Analogous to our above yarn-level modeling evaluations, we next assess shell fabrics in both equilibrium modeling and dynamic deformation exercises.

\paragraph*{Shell Fabric Equilibria}

In Figure~\ref{fig:pvc}, we simulate a $10 \times 10$~cm PVC sheet discretized with a 13k-triangle mesh and a thickness of 2.33mm. As in the yarn case, thickened shell models under standard barrier methods (again using Li et al.~\cite{Li2021CIPC}) suffer from contact locking, with contact forces improperly activated due to the fine discretization. This leads to artificial expansion and distorted equilibrium states. This issue becomes even more pronounced for thicker materials, as shown in Figure~\ref{fig:spandex-appendix} in Appendix~\ref{sec:appendix-additional-comparisons}, where a 5mm-thick fabric was simulated.
\input{images/tex/pvc_shell}

\paragraph*{Exercising Fabrics}
As with yarns, we next evaluate the shells under dynamic motions, comparing with the standard barrier formulation~\cite{Li2021CIPC}, as shown in Figure~\ref{fig:pvc}. Specifically, we simulate stretching and twisting of the same above PVC sheet. Due to element size falling below the activation threshold of the standard barrier method, contact forces similarly induce artificial expansion, resulting in a visibly inflated center region in the relaxed configuration. As in the yarn case, our barrier-filtering eliminates this artifact and produces undistorted equilibrium shapes. Following the equilibrium case, increasing material thickness (see Figure~\ref{fig:spandex-appendix} in Appendix~\ref{sec:appendix-additional-comparisons}) exacerbates these issues for the barrier method, producing spurious wrinkle patterns under stretching and shearing, and even generating a complete blow-up under twisting. Our \Method approach, in contrast, remains robust across all scenarios, ensuring stability while producing undistorted, contact-locking–free shell deformations.

\input{images/tex/cloth_on_corners}
\paragraph*{Cloth Drops}
In Figure~\ref{fig:cloth_on_corners}, we simulate a $20 \times 20$cm cloth sheet with a thickness of 0.7mm draping under gravity, with two corners pinned. The material parameters are set to a Young’s modulus of 0.8MPa, Poisson’s ratio of 0.3, and density of 500kg/m\textsuperscript{3}. The sheet is discretized using a $200 \times 200$ regular grid (40k triangles). As shown, a standard barrier method~\cite{Li2021CIPC} exhibits improper activation of contact forces, again resulting in noticeable expansion of the sheet.

In Figure~\ref{fig:cloth_on_sphere}, we simulate the same cloth material draping over a rigid sphere, now with 1mm thickness and discretized on a $100 \times 100$ regular grid (10k triangles). To consider the effect of standard shell-modeling practices to mitigate \emph{membrane locking}, we now additionally apply strain limiting\ \cite{Li2021CIPC}. However, the standard barrier method~\cite{Li2021CIPC} prematurely activates contact forces at the start of the simulation, causing artificial expansion. Strain-limiting barriers opposing this expansion then lock the configuration further, and result in extreme contact locking artifacts, with Schwarz–Lantern-like patterns as we see in Figure~\ref{fig:cloth_on_sphere}.
\input{images/tex/cloth_on_drake}
In contrast, our \Method approach avoids these issues and produces natural, physically plausible wrinkle patterns even with strain limiting enabled. 

Finally, in Figure~\ref{fig:cloth_on_drake} we further consider the generality of our barrier-filter method to enable simulation across resolutions, including graded and irregular meshings. Here we simulate a $1\times1$ meter cloth with $0.1$mm thickness. In this example, we use an extremely graded mesh, with non-uniform triangulation: higher resolution near the center and progressively coarser elements increasingly large toward the boundaries. Despite the central region containing elements well below the activation threshold of the standard barrier method, our \Method formulation captures wrinkles and draping behavior across the extreme resolution variations. 
\input{images/tex/drape_on_sphere}

%% file: images/tex/real_world_yarn_comp_closeup.tex
\begin{figure}[h]
    \centering
    \begin{tikzpicture}
    \node[anchor=south west,inner sep=0] (image) at (0,0){\includegraphics[width=.75\linewidth]{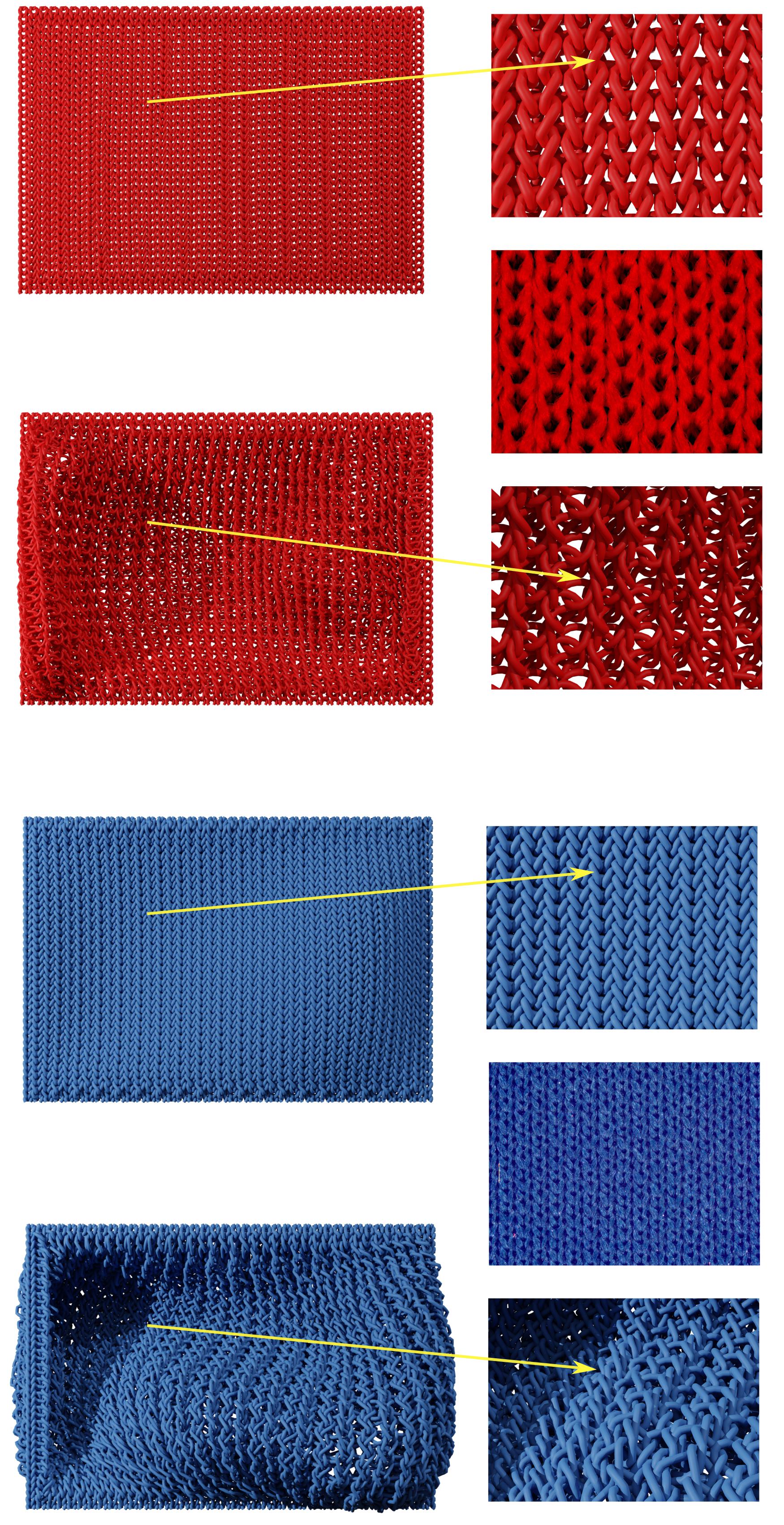}};
    \begin{scope}[x={(image.south east)},y={(image.north west)}]

        \node[anchor=south] at (0.42, 0.5) {A1 Pattern};
        \node[anchor=north] at (0.42, 0) {DKP Pattern};
    
        \node[anchor=south, rotate=90] at (0,0.91) {{Our Method}};
        \node[anchor=south, rotate=90] at (0.63,0.767) {{Real Image}};
        \node[anchor=south, rotate=90] at (0.0,0.63) {{Barrier}};

         \node[anchor=south, rotate=90] at (0,0.38) {{Our Method}};
        \node[anchor=south, rotate=90] at (0.63,0.232) {{Real Image}};
        \node[anchor=south, rotate=90] at (0.0,0.1) {{Barrier}};
    \end{scope}
    \end{tikzpicture}
    \caption{\textbf{Yarn relaxation comparison with real-world captures.} We compare the relaxation simulation results of the A1 pattern with real-world images both published in Sperl et al.~\cite{Sperl2022}. Close-up insets highlight differences between the simulated and physical patterns. Our barrier-filtered method accurately reproduces the relaxed yarn configuration without introducing artificial expansion artifacts. In contrast, the barrier method~\cite{Li2021CIPC} introduces spurious contact forces that distort the overall pattern, resulting in noticeable deviations from the real-world reference pattern.}
    \label{fig:real_world_yarns_comp}
\end{figure}

%% file: images/tex/self_pushing_with_Kaldor.tex
\begin{figure}[h]
    \begin{tikzpicture}
    \node[anchor=south west,inner sep=0] (image) at (0,0){\includegraphics[width=\linewidth]{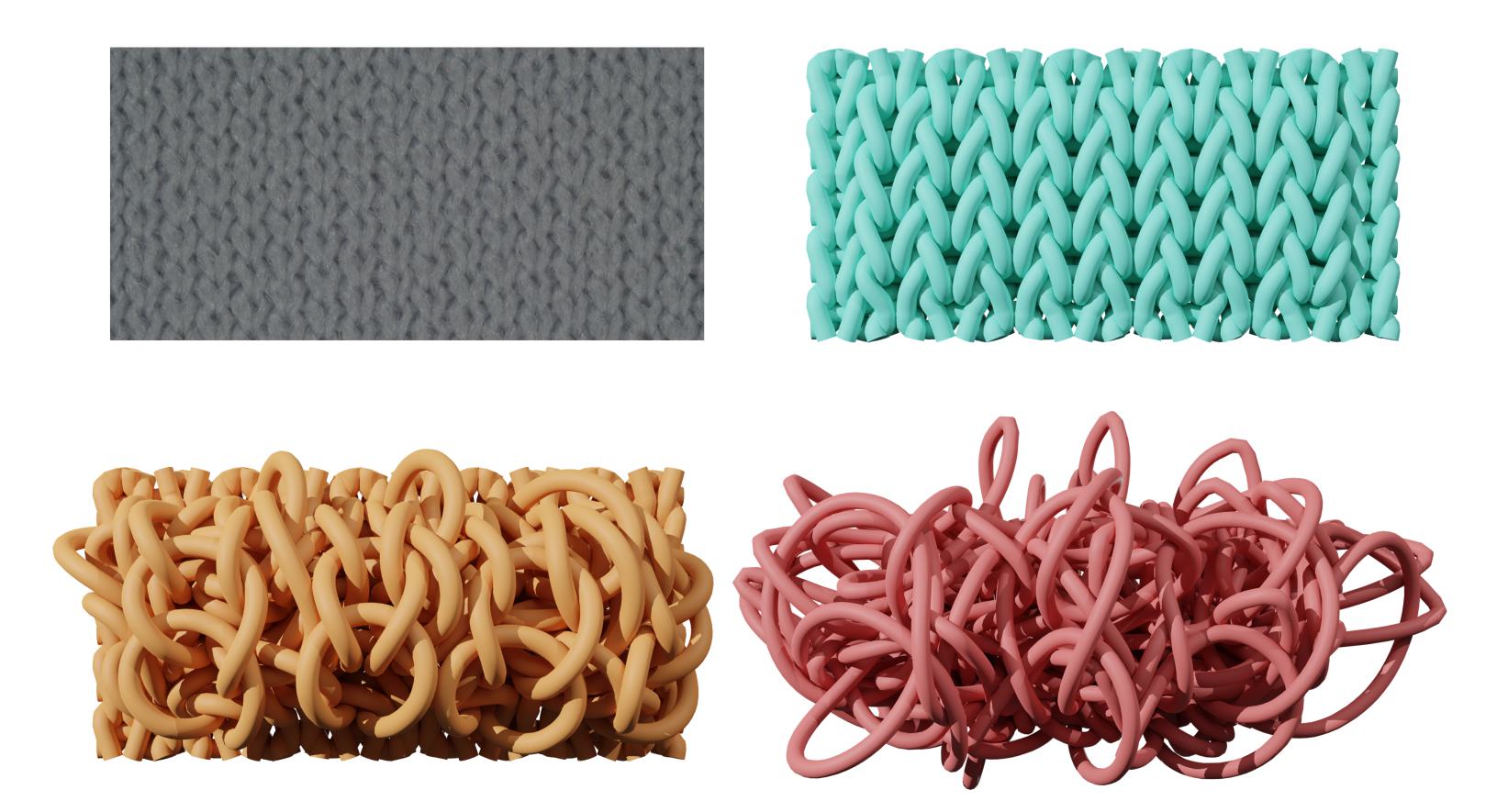}};
    \begin{scope}[x={(image.south east)},y={(image.north west)}]

        \node[anchor=north] at (0.27, 0.57) {\footnotesize{Real Image (DKIN1 Pattern)}};
        \node[anchor=north] at (0.75, 0.57) {\footnotesize{Our Method}};
        \node[anchor=north] at (0.25, 0.0) {\footnotesize{Culled}};
        \node[anchor=north] at (0.75, 0.0) {\footnotesize{Barrier}};

    \end{scope}
    \end{tikzpicture}
        \caption{\textbf{Yarn pattern relaxation method comparison with real-world capture.} We compare the simulated relaxation results with real-world images for the DKIN1 pattern from Sperl et al.~\cite{Sperl2022} with corresponding reported yarn material parameters, e.g., a yarn thickness of 0.181 mm. Here with an accurate discretization our barrier-filtered method closely reproduces the pattern structure. In contrast both the barrier method~\cite{Li2021CIPC} and the culled contact method~\cite{Kaldor2008} (the latter using a default of an aggressive culling radius of 11 elements) both suffer from contact-locking expansion artifacts.}
   \label{fig:toy_example_kaldor_self_pushing_forces}
\end{figure}

%% file: images/tex/twist_comp_yarn.tex
\begin{figure*}[htbp]
    \begin{tikzpicture}
    \node[anchor=south west,inner sep=0] (image) at (0,0){\includegraphics[width=\linewidth]{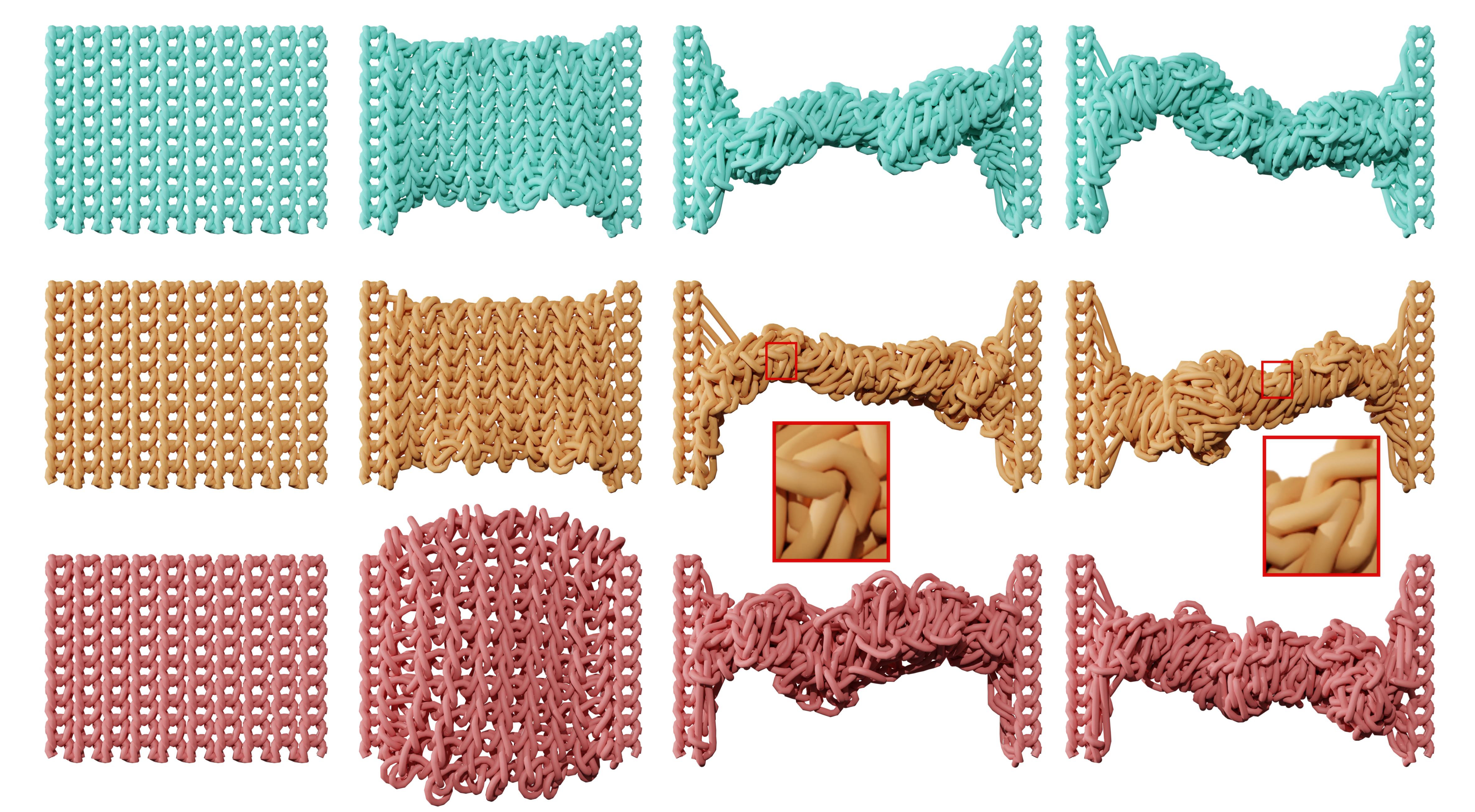}};
    \begin{scope}[x={(image.south east)},y={(image.north west)}]

        \node[anchor=south, rotate=90] at (0.02,0.85) {{Our Method}};
        \node[anchor=south, rotate=90] at (0.02,0.54) {{Culled}};
        \node[anchor=south, rotate=90] at (0.02,0.2) {{Barrier}};

        \node[anchor=north] at (0.13, 0) {Initial};
        \node[anchor=north] at (0.35, 0) {Equilibrium};
        \node[anchor=north] at (0.58, 0.0) {Mid};
        \node[anchor=north] at (0.85, 0.0) {Final};
    \end{scope}
    \end{tikzpicture}
   \caption{\textbf{Yarn twisting comparison with other methods.} We simulate the twisting of the A1 pattern~\cite{Sperl2022}, measuring $1.5\,\mathrm{cm} \times 1\,\mathrm{cm}$ with a thickness of $0.314\,\mathrm{mm}$. We show the initial input, the relaxed configuration, an intermediate frame during twisting, and the final twisted state. The barrier method~\cite{Li2021CIPC}, which does not cull contact pairs, already exhibits undesired expansion during the relaxation phase. The culling method introduced by Kaldor et al.~\cite{Kaldor2008} eliminates this initial expansion, but results in self-intersections due to over-culling. We visualize in zoom in to illustrate a few of the obvious, near-surface intersections, and note many intersections in the interior. In contrast, our filtered barrier approach resolves both issues simultaneously, preserving the correct topology while avoiding artificial deformation.}
    \Description{Yarn Twisting Comparisons}
    \label{fig:twisting_A1_small_comp_DF_CIPC_KALDOR_11}
\end{figure*}

%% file: images/tex/pvc_shell.tex
\begin{figure*}[htbp]
    \centering
    \begin{tikzpicture}
    \node[anchor=south west,inner sep=0] (image) at (0,0){\includegraphics[width=\linewidth]{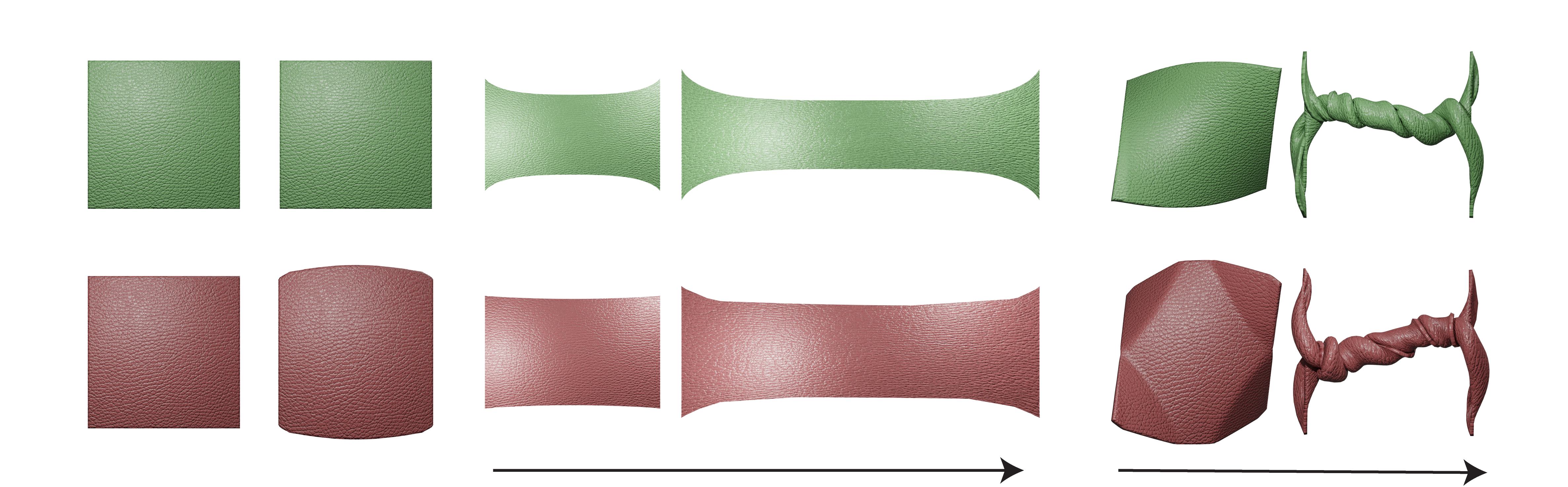}};
    \begin{scope}[x={(image.south east)},y={(image.north west)}]
        \node[anchor=south] at (0.105, 0.9) {Initial};
        \node[anchor=south] at (0.225, 0.89) {Equilibrium};
        \node[anchor=south] at (0.365, 0.9) {Mid};
        \node[anchor=south] at (0.54, 0.9) {Final};
        \node[anchor=south] at (0.76, 0.9) {Mid};
        \node[anchor=south] at (0.875, 0.9) {Final};
        \node[anchor=south] at (0.55, 0.02) {Stretching};
        \node[anchor=south] at (0.87, 0.02) {Twisting};

        \node[anchor=south, rotate=90] at (0.05, 0.26) {Barrier};
        \node[anchor=south, rotate=90] at (0.05, 0.71) {Our Method};
    \end{scope}
    \end{tikzpicture}
    \caption{\textbf{Stretching and twisting a pvc sheet.} We simulate the deformation of a $10 \times 10$cm pvc sheet with 2.23mm thickness, discretized using a 13k-triangle mesh. Due to the element size falling below the activation threshold, the standard barrier method~\cite{Li2021CIPC} activates contact forces prematurely, causing the sheet to expand artificially. In contrast, our \Method approach filters out these spurious contact forces, preserves the intended rest shape, and yields more accurate and physically plausible results.}
    \Description{pvc Shell Examples}
    \label{fig:pvc}
\end{figure*}

%% file: images/tex/cloth_on_corners.tex
\begin{figure}[htbp]
    \centering
    \begin{tikzpicture}
    \node[anchor=south west,inner sep=0] (image) at (0,0){\includegraphics[width=0.95\linewidth]{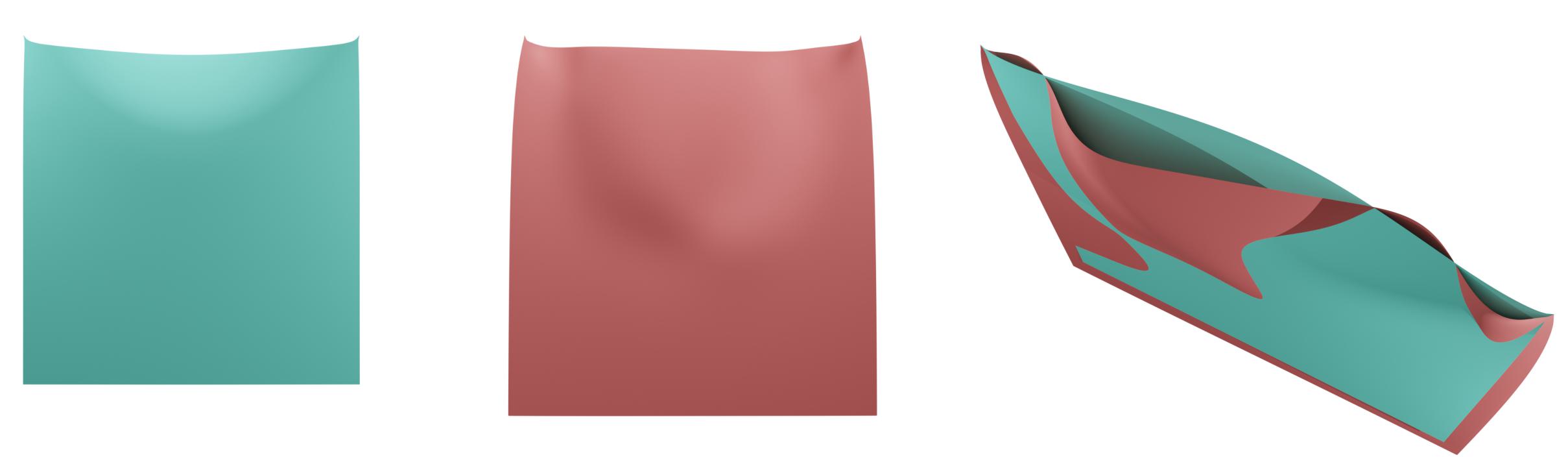}};
    \begin{scope}[x={(image.south east)},y={(image.north west)}]
        \node[anchor=north] at (0.12, 0) {Our Method};
        \node[anchor=north] at (0.45, 0) {Barrier};
        \node[anchor=north] at (0.8, 0) {Overlaid Top View};
    \end{scope}
    \end{tikzpicture}
    \caption{\textbf{Pinned cloth.} We simulate a $20 \times 20$cm cloth draping under gravity with two corners pinned. The sheet has a thickness of 0.7mm, Young’s modulus of 0.8MPa, and Poisson’s ratio of 0.3. It is discretized using a $200 \times 200$ regular grid (40k triangles). In both the overlaid view (right column), and side-by-side we see the standard barrier method's \cite{Li2021CIPC} significant expansion artifacts.}
    \Description{Cloth on drake}
    \label{fig:cloth_on_corners}
  \end{figure}

%% file: images/tex/cloth_on_drake.tex
\begin{figure}[htbp]
    \centering
    \begin{tikzpicture}
    \node[anchor=south west,inner sep=0] (image) at (0,0){\includegraphics[width=0.95\linewidth]{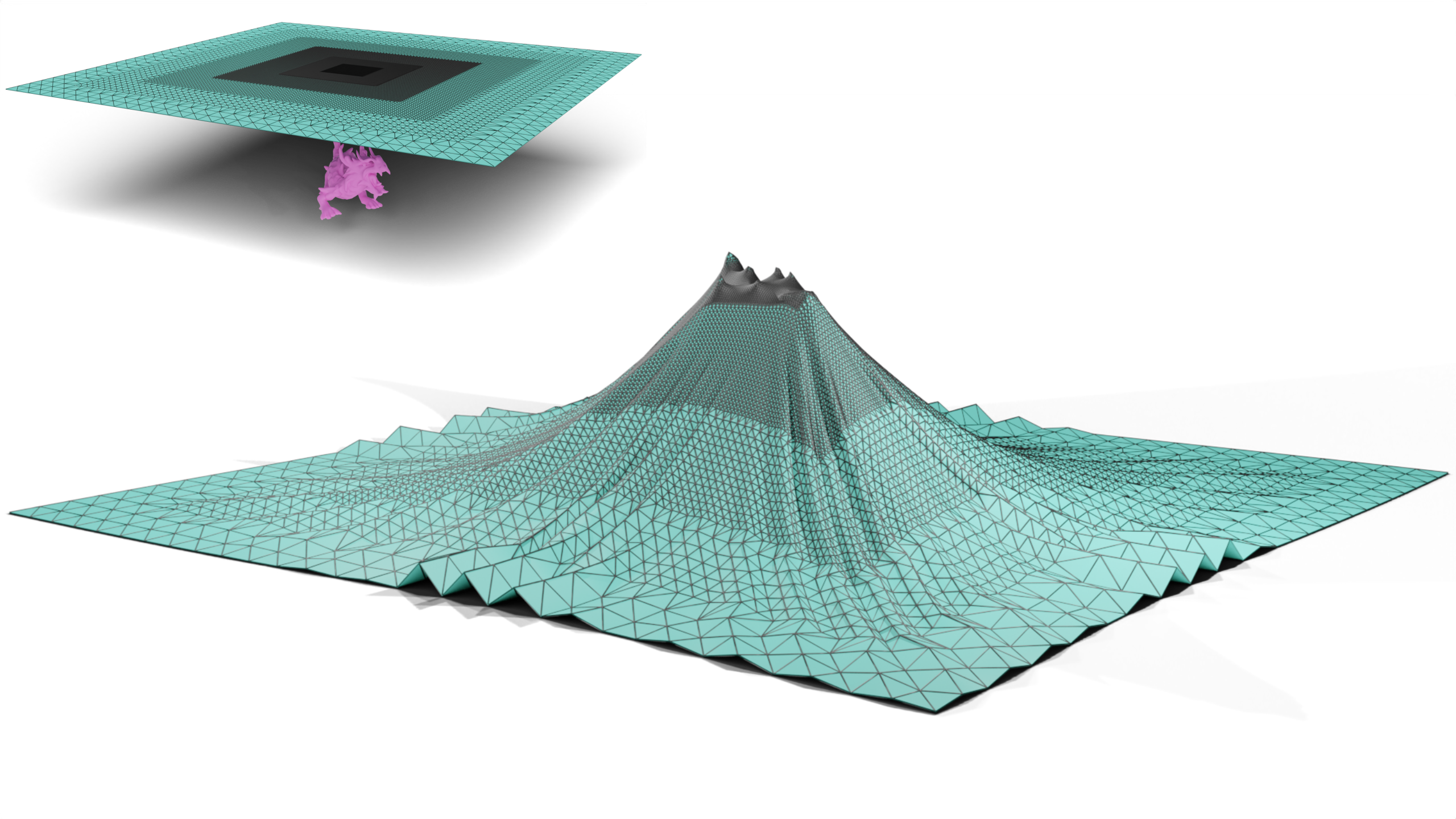}};
    \begin{scope}[x={(image.south east)},y={(image.north west)}]
        \node[anchor=north] at (0.2, 0.7) {Initial};
        \node[anchor=north] at (0.5, 0.1) {Final};
    \end{scope}
    \end{tikzpicture}
    \caption{\textbf{Vary mesh resolution.} Illustrating our barrier-filtering method's ability to safely resolve collisions with thickness across meshing variations, we drop a $1 \times 1$m cloth sheet (0.1 mm thickness), with extreme meshing variations (reaching, at center, 0.01~mm edge lengths), onto a sharp dragon geometry. Here, our barrier-filtering, avoids contact locking and resolves the shell simulation across the entire, resolution-varying domain without intersection.}
    \Description{Cloth on drake}
    \label{fig:cloth_on_drake}
  \end{figure}

%% file: images/tex/drape_on_sphere.tex
\begin{figure}[htbp]
    \centering
    \begin{tikzpicture}
    \node[anchor=south west,inner sep=0] (image) at (0,0){\includegraphics[width=0.8\linewidth]{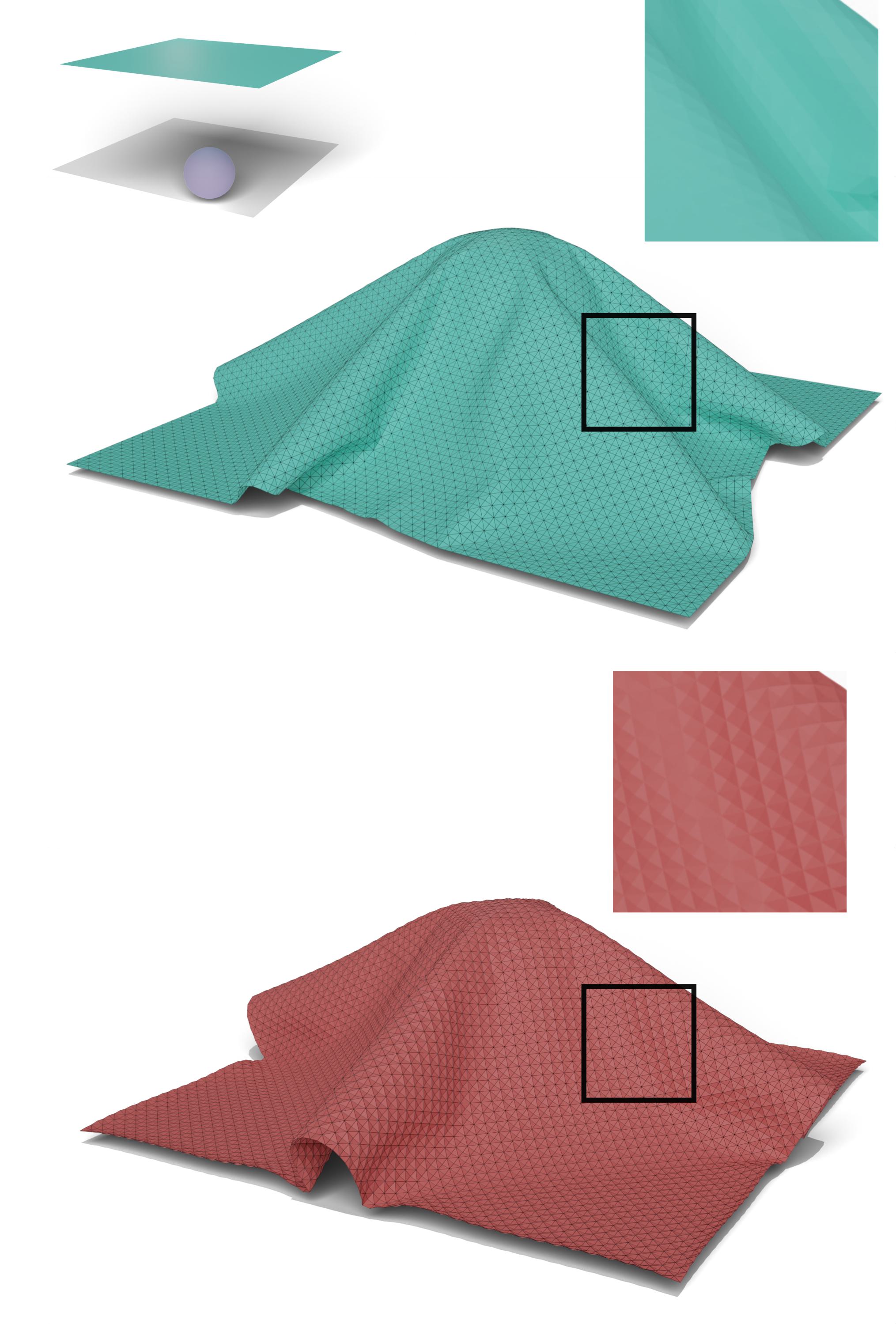}};
    \begin{scope}[x={(image.south east)},y={(image.north west)}]
        \node[anchor=north] at (0.2, 0.83) {{Initial}};
        \node[anchor=north] at (0.5, 0.55) {Our Method};
        \node[anchor=north] at (0.5, 0.05) {Barrier};
    \end{scope}
    \end{tikzpicture}
    \caption{\textbf{Thin shell dropped onto a sphere.} A $20 \times 20$cm sheet with 1 mm thickness is dropped onto a rigid sphere with strain-limiting enabled. The standard barrier method~\cite{Li2021CIPC} exhibits contact locking, resulting in the well-known Schwarz–Lantern-type artifacts. In contrast, our \Method method avoids contact-locking and producing material-consistent wrinkling and drape.}
    \Description{Cloth on sphere}
    \label{fig:cloth_on_sphere}
\end{figure}

%% file: 05-conclusion.tex
\section{Discussion, Limitations and Conclusion}

We have demonstrated and analyzed contact-locking artifacts in the modeling of yarns and thickened fabrics, a fundamental restriction blocking the reliable application of codimensional models for the accurate physical simulation of real-world fabric materials. We have then proposed a new and exceedingly simple, barrier-filtering contact-processing method for thickened codomain simulation of rods and shells that removes these restrictions, while guaranteeing non-intersecting yarn and fabric simulations, both for statics and dynamics modeling. Looking ahead there are limitations to address and interesting extensions to explore. First and foremost, heterogeneous materials with locally varying thickness properties, e.g. large ridging in corduroy wales, are an important category of thick materials generally suitable for codimensional modeling, that we do not yet well-resolve here. In theory, our current barrier-filtering strategy could be applied to such materials. However, for resolution of  surface interactions, a more localized filtering model, with changing activation distance filter updates per stencil, would likely be a significantly better approach in such contexts for capturing varying anisotropic contact responses. Likewise, we observe that, in all such barrier-based models, the transverse compression response modeled by the contact barrier is largely decoupled from the remaining constitutive material model. Future development of barrier-filtering that jointly accounts for coupling with stretch (axial and membrane) and bending response should be an exciting avenue of future exploration.

%% file: 0A-implementation-details_appendix.tex
\section{Implementation Details}
\label{sec:appendix-implementation}
We implement all methods in a common C++ test harness, applying CHOLMOD~\cite{Chen2008Cholmod}, for linear solves and Eigen for remaining linear algebra routines~\cite{eigenweb}. For robust line-search filtering we evaluate continuous
collision detection queries with a spatial-hash culled ACCD~\cite{Li2021CIPC}. Fabrics and yarns in all experiments are consistently implemented with Neo-Hookean membrane~\cite{vougalibshell} and discrete hinge bending~\cite{Grinspun2003, Tamstorf2013}  for shell elastics, Discrete Rods stretch and bending models~\cite{Bergou2008} for rod elastics, IPC~\cite{Li2020IPC} energies for contact and friction, and the C-IPC barrier~\cite{Li2021CIPC} for strain limiting. 

All simulations are performed with a fixed time step of $0.04$s, using implicit Euler integration. Within each time step, we employ the barrier-aware projected Newton solver~\cite{Li2020IPC}, terminating when the Newton decrement falls below $10^{-4}$ for yarn-level examples and $10^{-3}$ for fabric-level examples.
In examples comparing with prior culled and barrier methods we use a biphasic thickness of $\eta = 0.9 \cdot \min(h, \bar{d}_{\min})$. This is necessary as these methods would otherwise, with larger $\eta$ fail  at start of simulation solve. In contrast, as we demonstrate in Figure~\ref{fig:eta_comp}, our barrier-filtered method allows setting  biphasic $\eta$ anywhere within the material thickness range (i.e., $\eta \in [0,h]$) and so allows for a full range of compressive behavior. For other non-comparison examples in Figures~\ref{fig:cloth_on_drake} and~\ref{fig:eta_comp} we use $\eta=0.9 \cdot h$.
All culling examples, following Kaldor\ \cite{Kaldor2008}, apply the default culling radius of 11, with the exception of our knot test where we apply a smallest possible radius of 5 that avoids contact-locking for the culled simulation in those tests.  
For all fabric simulations, we use a Young’s modulus of $0.8$MPa, Poisson’s ratio of $0.3$, and density of $500$kg/m\textsuperscript{3}, please refer to Table~\ref{tab:materials-cloth-appendix} for more details. For yarn pattern material information, please refer to Table~\ref{tab:materials-appendix} for details. As expected, given the same underlying simulation test-harness, we observe no significant slowdown nor speedup in comparing across simulations between our barrier-filtering, culling\ \cite{Kaldor2008} and baseline barrier\ \cite{Li2021CIPC} methods, emphasizing that our method's improvements do not require extra simulation costs.

\input{images/tex/table_summary_materials_cloth_appendix}
\input{images/tex/table_summary_materials_appendix}
\input{images/tex/helix}
\input{images/tex/eta_comp}

%% file: images/tex/table_summary_materials_cloth_appendix.tex
\begin{table}[!h]
\vspace{1em} %
\centering
\footnotesize
{
\begin{tabular}{|l|l|l|l|l|}
\hline
\textbf{Description} & \textbf{Pattern Size} & \textbf{\#Triangles} & \textbf{Thickness} \\
\hline
Fig.~\ref{fig:pvc} & $10\times10$~cm & 13k & 2.23mm \\
Fig.~\ref{fig:cloth_on_corners} & $20 \times 20$~cm & 40k & 0.7mm \\
Fig.~\ref{fig:cloth_on_drake} & $1 \times 1$~m & 100k & 0.1mm \\
Fig.~\ref{fig:cloth_on_sphere} & $20 \times 20$~cm & 10k & 1mm \\
Fig.~\ref{fig:spandex-appendix}  & $10\times10$~cm & 13k & 5mm \\
Fig.~\ref{fig:eta_comp} & $20 \times 20$~cm & 10k & 5mm \\
\hline
\end{tabular}
}
\caption{Pattern information for all cloth figures in the paper. For all cloth, we use isotropic material with Young's Modulus = 0.8MPa, Poisson's ratio = 0.3, and density = $500\text{kg}/\text{m}^3$.}
\label{tab:materials-cloth-appendix}
\end{table}

%% file: images/tex/table_summary_materials_appendix.tex
\begin{table}[htb]
\vspace{1em} %
\centering
\footnotesize
{

\begin{tabular}{|l|l|l|l|l|}
\hline
\textbf{Description} & \textbf{Material} & \textbf{Pattern Size} & \textbf{\#Segs} & \textbf{Thickness} \\
\hline
Fig.~\ref{fig:resolution_decimation_example} top & 150D/72F & $3.5 \times 2.5$~mm & 1800 & 0.177mm \\
Fig.~\ref{fig:resolution_decimation_example} mid & 150D/72F & $3.5 \times 2.5$~mm & 800 & 0.177mm \\
Fig.~\ref{fig:resolution_decimation_example} bottom & 150D/72F  & $3.5 \times 2.5$~mm & 500 & 0.177mm \\
Fig.~\ref{fig:helix_pull_through} & 75D/96F & $5$~mm & 18 & 0.5mm \\
Fig.~\ref{fig:knots_pull_through_example} top & 75D/96F  & $5$~mm & 30 & 0.3mm \\
Fig.~\ref{fig:knots_pull_through_example} middle & 75D/96F  & $5$~mm & 60 & 0.3mm \\
Fig.~\ref{fig:knots_pull_through_example} bottom & 75D/96F  & $5$~mm & 120 & 0.3mm \\
Fig.~\ref{fig:toy_example_kaldor_self_pushing_forces} & 167D/72F  & $5 \times 3$~mm & 20000 & 0.181mm \\
Fig.~\ref{fig:real_world_yarns_comp},~\ref{fig:yarn-level-equilibria}$^\text{A1}$ & 75D/72F  & $5 \times 4$~cm & 62000 & 0.314mm \\
Fig.~\ref{fig:real_world_yarns_comp},~\ref{fig:yarn-level-equilibria}$^\text{DKP}{}$ & 75D/72F  & $2 \times 1.5$~cm & 80000 & 0.177mm \\
Fig.~\ref{fig:our_barrier_comparison},~\ref{fig:slow_twist_A1} & 75D/72F  & $5 \times 4$~cm & 62000 & 0.314mm \\
Fig.~\ref{fig:twisting_A1_small_comp_DF_CIPC_KALDOR_11} & 75D/72F  & $1.5 \times 1$~cm & 5000 & 0.314mm \\
\hline
\end{tabular}
}
\caption{Material parameters and pattern information for all yarn figures in the paper.}
\label{tab:materials-appendix}
\end{table}

%% file: images/tex/helix.tex
\begin{figure}[!t]
    \begin{tikzpicture}
    \node[anchor=south west,inner sep=0] (image) at (0,0){\includegraphics[width=0.9\linewidth]{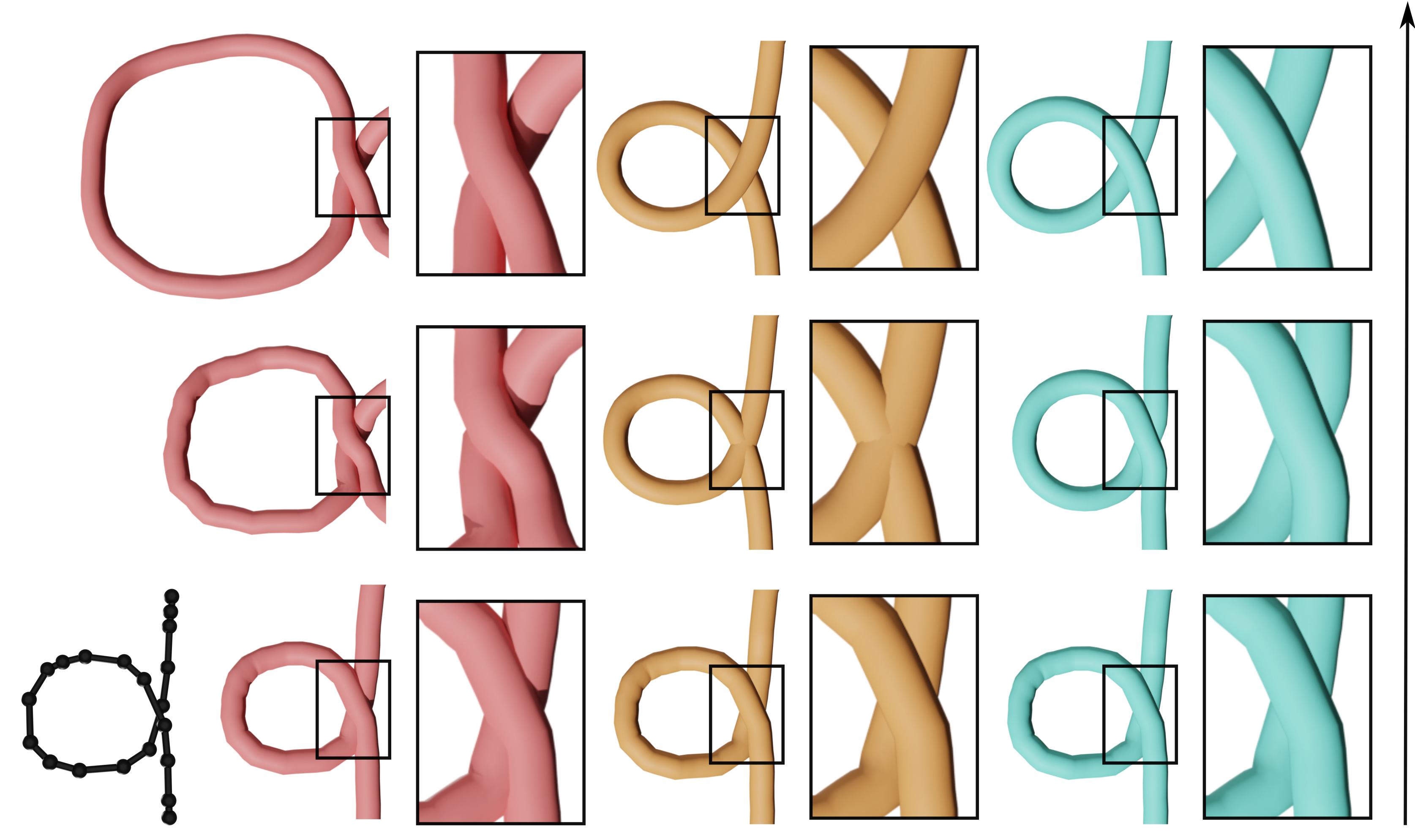}};
    \begin{scope}[x={(image.south east)},y={(image.north west)}]
        \node[anchor=north] at (0.1, 0) {Polylines};
        \node[anchor=north] at (0.3, 0) {Barrier};
        \node[anchor=north] at (0.6, 0) {Culled};
        \node[anchor=north] at (0.85, 0) {Our Method};
        \node[anchor=north, rotate=90] at (1, 0.5) {Simulating};
    \end{scope}
    \end{tikzpicture}
   \caption{\textbf{Loop test.} Starting from the input configuration (rendered on bottom, including polyline view) we simulate a 0.5mm thick yarn to equilibrium, showing progressive simulation steps increasing in time vertically. \textbf{\textcolor{yarnRed}{left}}: Barrier-method forces \cite{Li2021CIPC} preserve the loops topology but immediately begin activating nonphysical contact forces, leading to artificially expanded loop.  \textbf{\textcolor{yarnOrange}{middle}}: Culling \cite{Kaldor2008} avoids the expansion, but introduces self-intersection (highlighted in the zoomed-ins) and pull through. \textbf{\textcolor{yarnGreen}{right}}: in contrast, our Barrier-Filtering finds a smooth, unexpanded loop without pull-through.}
  \label{fig:helix_pull_through}
\end{figure}

%% file: images/tex/eta_comp.tex
\begin{figure*}[htbp]    
    \centering
    \begin{tikzpicture}
    \node[anchor=south west,inner sep=0] (image) at (0,0) {\includegraphics[width=\linewidth]{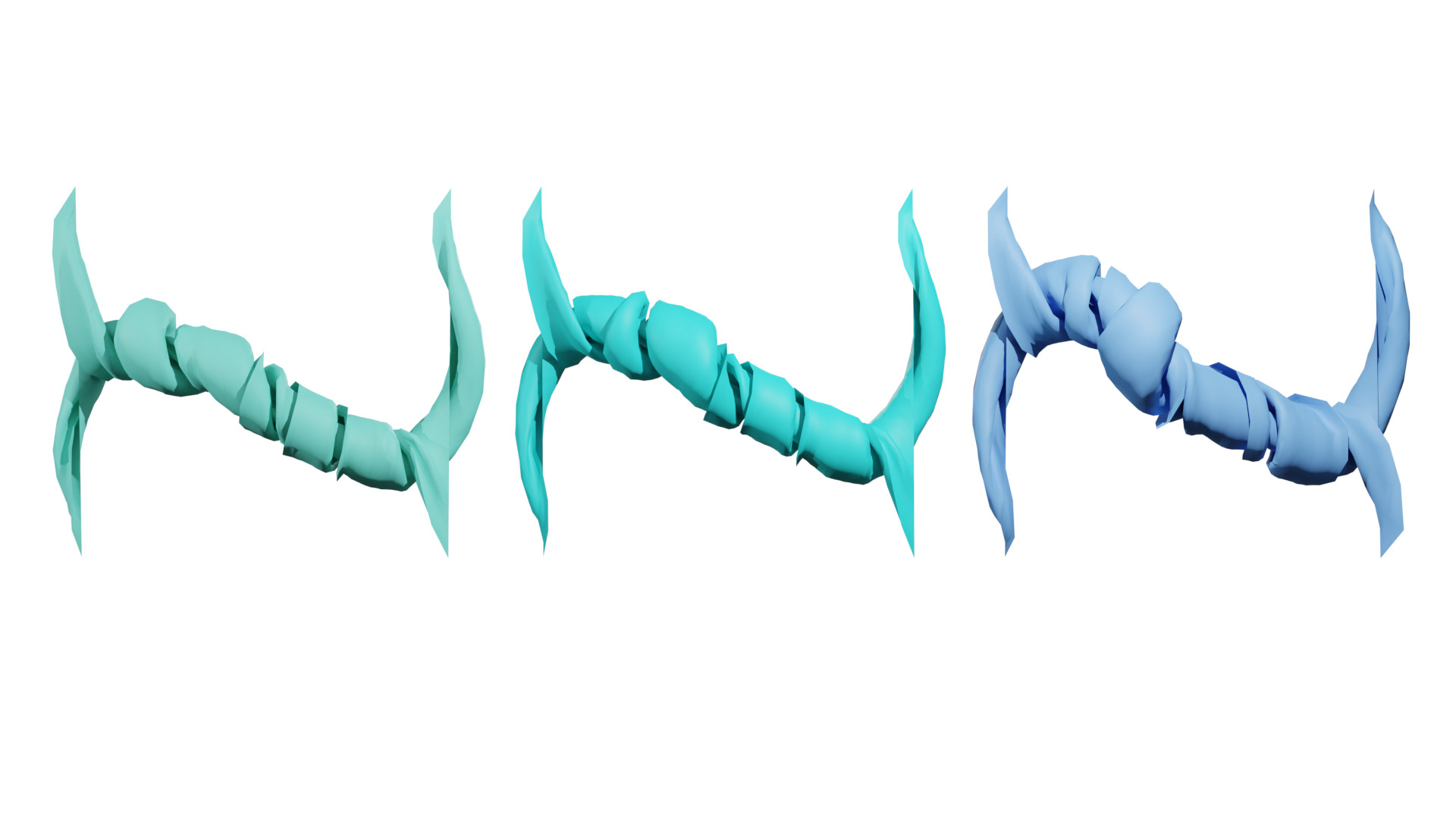}};
    \begin{scope}[x={(image.south east)},y={(image.north west)}]
        \node[anchor=north] at (0.18, 0.2) {$\eta = \bar{d}_{min}$};
        \node[anchor=north] at (0.5, 0.2) {$\eta = \frac{h}{2}$};
        \node[anchor=north] at (0.82, 0.2) {$\eta = h$};
    \end{scope}
    \end{tikzpicture}
    \caption{\textbf{$\eta$ effect on simulation.} We twist a $20 \times 20$cm cloth choosing 3 different values for $\eta$. The sheet has a thickness ($h$) of 5mm, Young’s modulus of 0.8MPa, and Poisson’s ratio of 0.3. It is discretized using a $100 \times 100$ regular grid (10k triangles). Different $\eta$ values lead to different states of the twisting simulation, our filtered barrier allows to set the value that better fits the desired biphasic behavior of the real-world material being simulated, as in \cite{Li2021CIPC} but without resolution restrictions, starting from $\eta = \bar{d}_{min}$ (left) increasing the value through $\eta = \frac{h}{2}$ (center) up to $\eta = h$ (right).}
    \Description{Eta value effect on simulation}
    \label{fig:eta_comp}
  \end{figure*}

%% file: 0B-more-comparisons_appendix.tex
\section{Additional Comparison Examples}
\label{sec:appendix-additional-comparisons}
In this Section we provide additional comparisons with state-of-the-art contact handling alternatives.

\subsection{Pull Through: Knot Test}
In the \emph{knot} test we use real-world yarn parameters from the dataset of Sperl et al.\ \cite{Sperl2022} (75D/72F Polyester, 0.5 mm thickness) to simulate pulling both ends of a knotted yarn curve at three increasing resolutions of 30, 60, and 120 segments. As illustrated in Figure~\ref{fig:knots_pull_through_example}, culling methods~\cite{Kaldor2008} fail in all cases, allowing the yarn to pass through itself and ultimately losing the knot structure. In contrast, our method consistently prevents self-intersections and preserves the knot topology across all tested resolutions. 

\subsection{Loop Test.}  
Figure~\ref{fig:helix_pull_through} illustrates an \textit{extreme} didactic example where we show that fixed window-sized culling is inappropriate for robust yarn-level  simulation. Here (again using the 75D/72F Polyester, 0.5 mm thickness yarn as above) the barrier method~\cite{Li2021CIPC} successfully prevents self-intersection in the loop, but introduces significant shape distortion. When we set the culling method's~\cite{Kaldor2008} culling radius to its default large window of 11 elements, we see that it preserves the intended curve shape but permits self-intersections. In contrast our filtering obtains a smooth, unexpanded loop, without pull-through, nor need to find a per-example appropriate culling radius.

\subsection{Yarn Pattern Examples.}  
Figures~\ref{fig:yarn-level-equilibria} and~\ref{fig:our_barrier_comparison} present additional yarn pattern comparisons between the barrier method~\cite{Li2021CIPC} and our approach. The barrier-based method suffers from self-expansion artifacts due to contact locking, whereas our method fully eliminates these artifacts and produces stable, artifact-free equilibria. Also note Figure~\ref{fig:slow_twist_A1}, where high-speed twisting yarn deformations and compression shown in Figure~\ref{fig:our_barrier_comparison} is contrasted with a slower motion twisting exercise.

\subsection{Shell Fabrics with Increased Thickness.}  
We further simulate a fabric with the same size and discretization as the PVC material, but with a greater thickness (5mm). As shown in Figure~\ref{fig:spandex-appendix}, increasing the thickness exacerbates contact locking, resulting in severe self-pushing artifacts. Our method robustly avoids these issues, producing consistent and realistic deformations.

\subsection{Parameter $\eta$ value comparison.}
In Figure~\ref{fig:eta_comp} we twist a fabric using 3 different values for $\eta$ with our filtered-barrier method. We show how our method enables application of target $\eta$ values independent of desired simulation resolution and so enables modeling different geometric thickness behaviors in the twisting simulation.

\input{images/tex/yarn_level_equilibria.tex}
\input{images/tex/ours_barrier_comparison.tex}

\input{images/tex/spandex_shell_appendix}
\input{images/tex/slow_twist_yarn}

%% file: images/tex/yarn_level_equilibria.tex
\begin{figure*}[h]
    \begin{tikzpicture}
    \node[anchor=south west,inner sep=0] (image) at (0,0){\includegraphics[width=.9\textwidth]{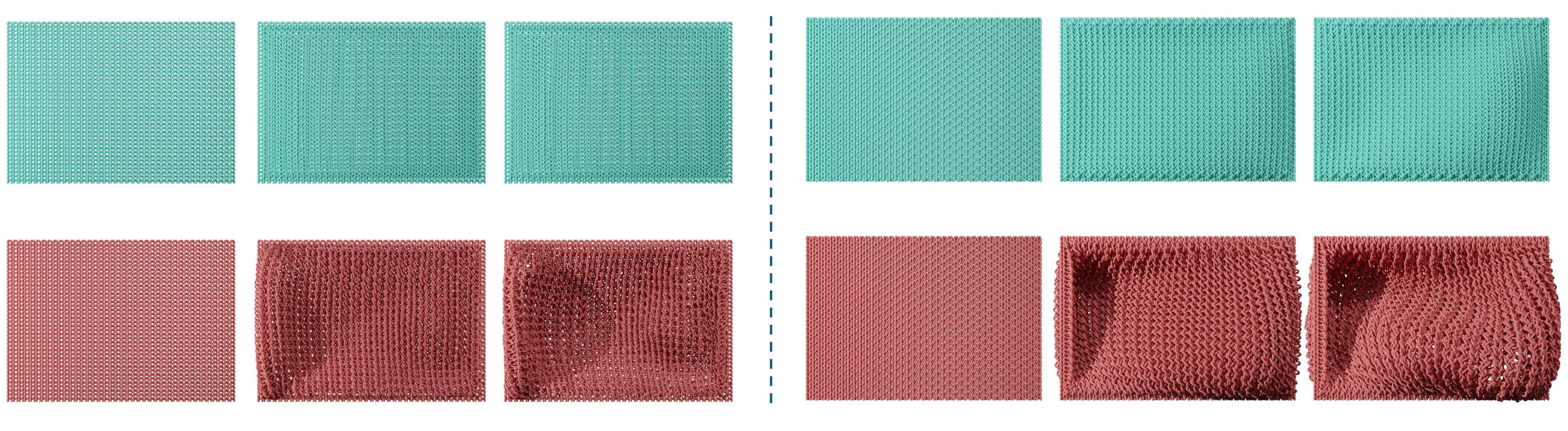}};
    \begin{scope}[x={(image.south east)},y={(image.north west)}]
    
        \node[anchor=south, rotate=90] at (0,0.75) {Our method};
        \node[anchor=south, rotate=90] at (0.0,0.25) {Barrier};

        \node[anchor=north] at (0.075,0) {Input};
        \node[anchor=north] at (0.23,0) {Mid};
        \node[anchor=north] at (0.39,0) {Equilibrium};
        \node[anchor=north] at (0.59,0) {Input};
        \node[anchor=north] at (0.75,0) {Mid};
        \node[anchor=north] at (0.91,0) {Equilibrium};

        \node[anchor=south] at (0.25,1) {A1 Pattern Size: $5cm \times 4cm \times 0.314mm$};
        \node[anchor=south] at (0.75,1) {DKP Pattern Size: $2cm \times 1.5cm \times 0.177mm$};
    \end{scope}
    \end{tikzpicture}
    \caption{\textbf{Knitted pattern equilibria with real-world yarn materials.} We present two examples of yarn equilibrium patterns using materials from the dataset of Sperl et al.~\cite{Sperl2022} (A1 and DKP patterns respectively). For each material, we show the input pattern, an intermediate state, and the final relaxed configuration. The left example begins with a $5\,\mathrm{cm} \times 4\,\mathrm{cm}$ rectangle of $0.314\,\mathrm{mm}$ thickness, while the right starts with a $2\,\mathrm{cm} \times 1.5\,\mathrm{cm}$ rectangle of $0.177\,\mathrm{mm}$ thickness. As shown, our method successfully relaxes input patterns to their equilibrium states with minimal deformation, whereas the barrier method~\cite{Li2021CIPC} produces large-scale distortions from non-physical expansive contact forces.}
    \label{fig:yarn-level-equilibria}
\end{figure*}

%% file: images/tex/ours_barrier_comparison.tex
\begin{figure*}[h]
    \centering
    \begin{tikzpicture}
    \node[anchor=south west,inner sep=0] (image) at (0,0){\includegraphics[width=0.95\linewidth]{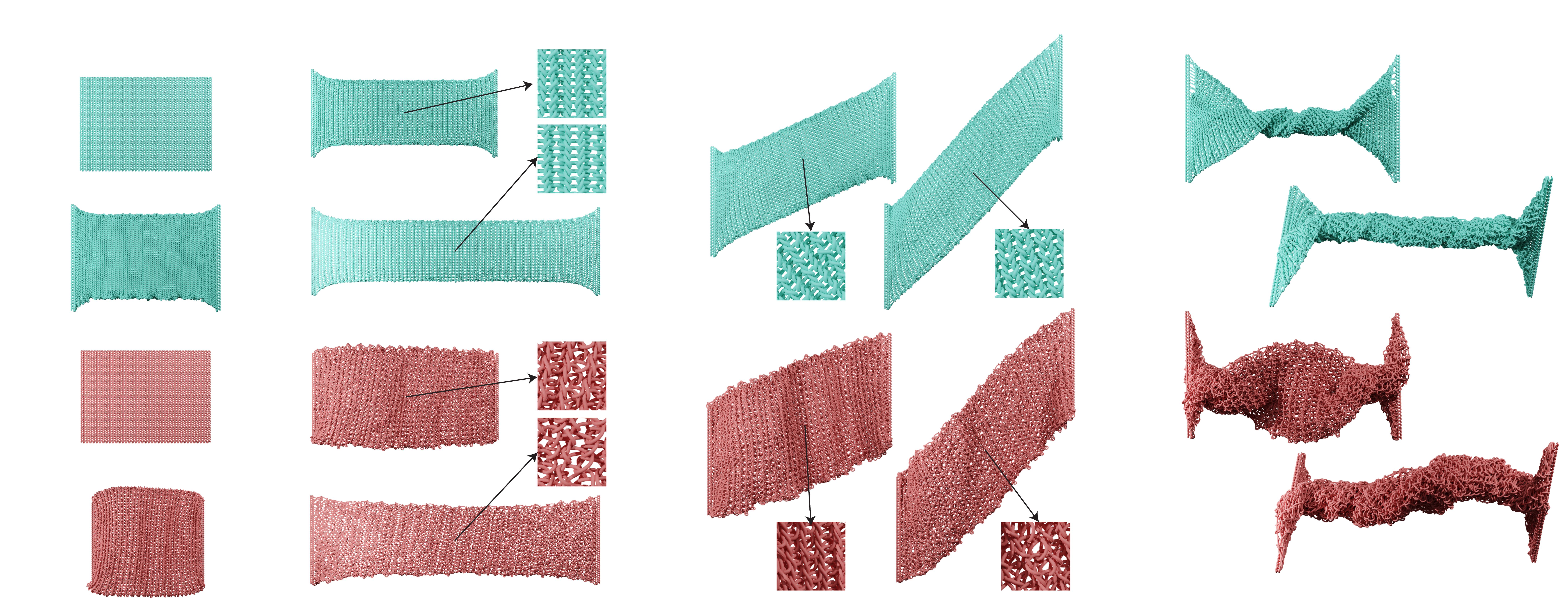}};
    \begin{scope}[x={(image.south east)},y={(image.north west)}]
        \node[anchor=south] at (0.095, 0.87) {\footnotesize{Initial}};
        \node[anchor=south] at (0.095, 0.64) {\footnotesize{Equilibrium}};
        
        \node[anchor=south] at (0.095, 0.41) {\footnotesize{Initial}};
        \node[anchor=south] at (0.095, 0.18) {\footnotesize{Equilibrium}};

        \node[anchor=south] at (0.3, 0.9) {Stretching};
        \node[anchor=south] at (0.55, 0.9) {Shearing};
        \node[anchor=south] at (0.85, 0.9) {Twisting};  

        \node[anchor=west] at (0.195, 0.9) {\footnotesize{Mid}};
        \node[anchor=west] at (0.195, 0.68) {\footnotesize{Final}};
        \node[anchor=west] at (0.195, 0.44) {\footnotesize{Mid}};
        \node[anchor=west] at (0.195, 0.2) {\footnotesize{Final}};

        \node[anchor=west] at (0.45, 0.8) {\footnotesize{Mid}};
        \node[anchor=west] at (0.6, 0.87) {\footnotesize{Final}};
        \node[anchor=west] at (0.45, 0.39) {\footnotesize{Mid}};
        \node[anchor=west] at (0.6, 0.425) {\footnotesize{Final}};

        \node[anchor=east] at (0.755, 0.82) {\footnotesize{Mid}};
        \node[anchor=east] at (0.815, 0.62) {\footnotesize{Final}};
        \node[anchor=east] at (0.755, 0.37) {\footnotesize{Mid}};
        \node[anchor=east] at (0.815, 0.17) {\footnotesize{Final}};
        
    \end{scope}
    \end{tikzpicture}
    
    \caption{\textbf{Exercising yarn patterns with stretching, shearing, and twisting}. 
    We compare top (green), our barrier-filtered method with bottom, barrier-based, unculled, simulation (red) \cite{Li2021CIPC} using the A1 fabric pattern with the 75D/72F Polyester (0.314 mm thickness) \cite{Sperl2022}. On the left, we first relax our initial yarn patterns to equilibrium without gravity. Here we already see significant expansion artifacts in the barrier simulations. Next we exercise this initial relaxed state with respectively a stretch, shear and twist test, showing both an intermediate simulation frame midway (\textit{Mid}) and the final state (\textit{Final}). For the mid-states, we observe that the contact-locking expansion again generates significant artifacts, while in final states, contact-locking artifacts are only excerbated by the more extreme pattern deformations.}
    \label{fig:our_barrier_comparison}
  \end{figure*}

%% file: images/tex/spandex_shell_appendix.tex
\begin{figure*}[htpb]
    \centering
    \begin{tikzpicture}
    \node[anchor=south west,inner sep=0] (image) at (0,0){\includegraphics[width=0.9\linewidth]{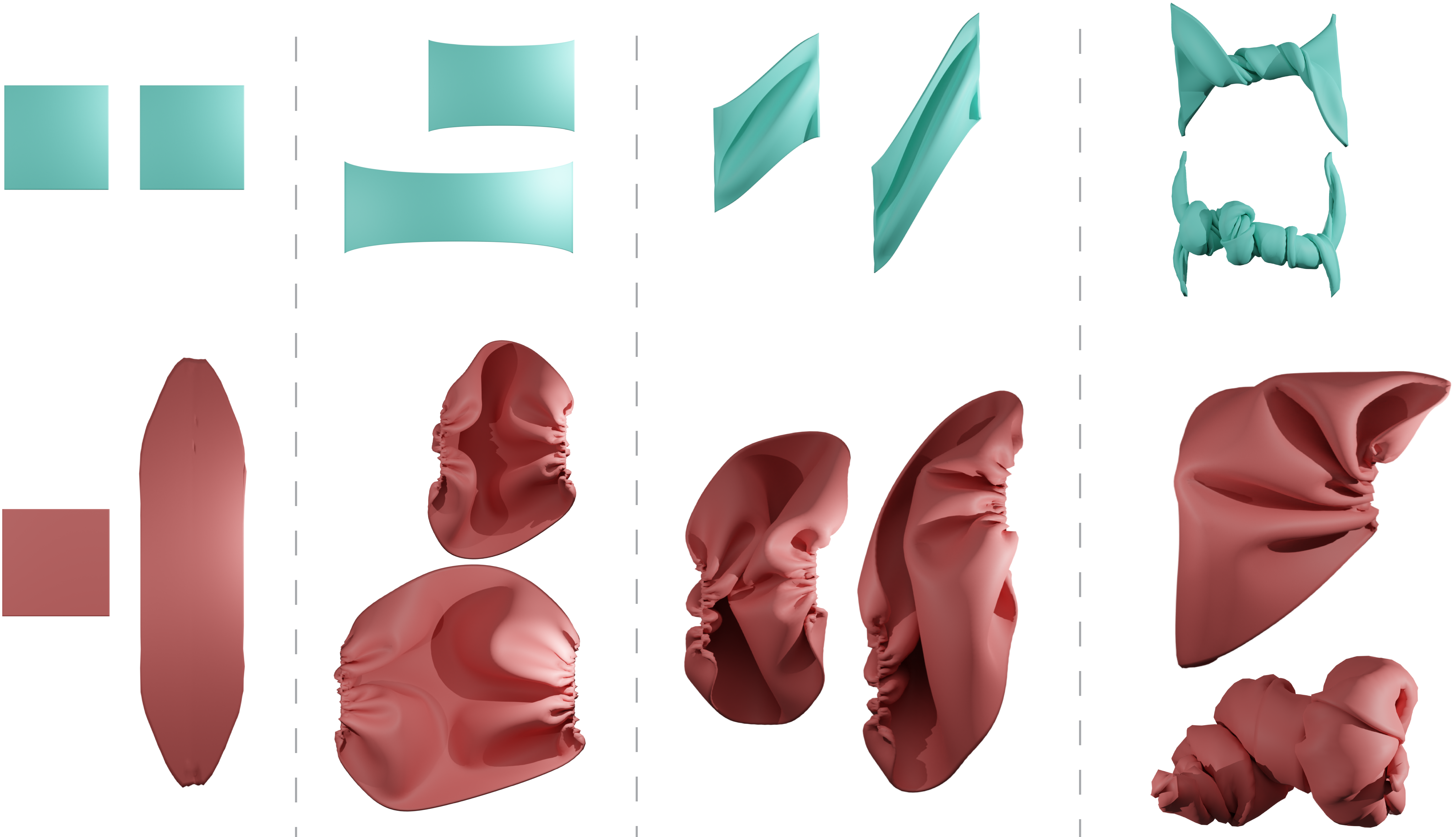}};
    \begin{scope}[x={(image.south east)},y={(image.north west)}]
        \node[anchor=south] at (0.03, 0.96) {Initial};
        \node[anchor=south] at (0.13, 0.95) {Equilibrium};
        \node[anchor=south] at (0.3, 0.96) {Stretching};
        \node[anchor=south] at (0.6, 0.96) {Shearing};
        \node[anchor=south] at (0.87, 0.96) {Twisting};
        \node[anchor=south, rotate=90] at (0, 0.33) {Barrier};
        \node[anchor=south, rotate=90] at (0, 0.83) {Our Method};

        \node[anchor=east] at (0.29, 0.9) {Mid};
        \node[anchor=north] at (0.32, 0.7) {Final};
        \node[anchor=east] at (0.29, 0.45) {Mid};
        \node[anchor=north] at (0.32, 0.03) {Final};
        
        \node[anchor=north] at (0.53, 0.78) {Mid};
        \node[anchor=north] at (0.65, 0.7) {Final};
        \node[anchor=north] at (0.53, 0.12) {Mid};
        \node[anchor=north] at (0.65, 0.03) {Final};
        
        \node[anchor=east] at (0.8, 0.9) {Mid};
        \node[anchor=east] at (0.8, 0.72) {Final};
        \node[anchor=east] at (0.8, 0.35) {Mid};
        \node[anchor=east] at (0.8, 0.1) {Final};
        
    \end{scope}
    \end{tikzpicture}
    \caption{\textbf{Stretching, shearing, and twisting a spandex sheet.} We perform the same thin shell simulation setup as in Figure~\ref{fig:pvc} (PVC sheet exercising), using identical dimensions and discretization, but with a thicker spandex material (5mm thickness). The contact-pushing artifacts become significantly more pronounced in this case, leading to highly distorted and physically implausible results under the standard barrier method\cite{Li2021CIPC}.}
    \Description{Spandex Shell Examples}
    \label{fig:spandex-appendix}
\end{figure*}

%% file: images/tex/slow_twist_yarn.tex
\begin{figure*}[t]
    \centering
    \includegraphics[width=0.6\linewidth]{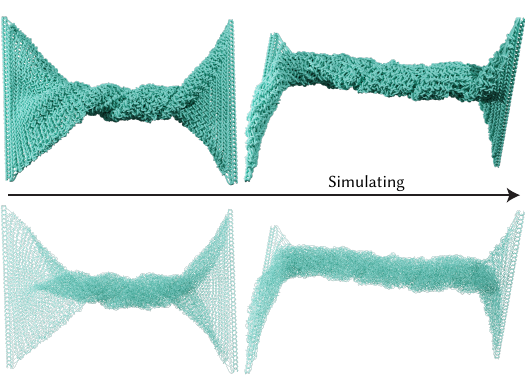}
    \caption{\textbf{Slow-Twisting}. 
We apply a less extreme, slower twisting test, using the same relaxed pattern from Figure~\ref{fig:our_barrier_comparison} to our method demonstrating (top) the more uniform pattern behavior, especially along rotated boundaries, as expected for this slower exercise. Bottom: we show the codimensional simulation's corresponding polyline discretization for reference. }
    \label{fig:slow_twist_A1}
\end{figure*}

%% file: main.bbl
\newcommand{\etalchar}[1]{$^{#1}$}
\begin{thebibliography}{\uppercase{CLMMO14}}

\bibitem[And24]{Ando2024}
\textsc{Ando R.}:
\newblock A cubic barrier with elasticity-inclusive dynamic stiffness.
\newblock \emph{ACM Trans. Graph. 43}, 6 (Nov. 2024).
\newblock URL: \url{https://doi.org/10.1145/3687908}, \href {https://doi.org/10.1145/3687908} {\path{doi:10.1145/3687908}}.

\bibitem[BAV{\etalchar{*}}10]{Bergou2010}
\textsc{Bergou M., Audoly B., Vouga E., Wardetzky M., Grinspun E.}:
\newblock Discrete viscous threads.
\newblock \emph{ACM Trans. Graph. 29}, 4 (July 2010).
\newblock URL: \url{https://doi.org/10.1145/1778765.1778853}, \href {https://doi.org/10.1145/1778765.1778853} {\path{doi:10.1145/1778765.1778853}}.

\bibitem[BW98]{Baraff1998}
\textsc{Baraff D., Witkin A.}:
\newblock Large steps in cloth simulation.
\newblock In \emph{Proceedings of the 25th Annual Conference on Computer Graphics and Interactive Techniques} (New York, NY, USA, 1998), SIGGRAPH '98, Association for Computing Machinery, p.~43–54.
\newblock URL: \url{https://doi.org/10.1145/280814.280821}, \href {https://doi.org/10.1145/280814.280821} {\path{doi:10.1145/280814.280821}}.

\bibitem[BWR{\etalchar{*}}08]{Bergou2008}
\textsc{Bergou M., Wardetzky M., Robinson S., Audoly B., Grinspun E.}:
\newblock Discrete elastic rods.
\newblock In \emph{ACM SIGGRAPH 2008 Papers} (New York, NY, USA, 2008), SIGGRAPH '08, Association for Computing Machinery.
\newblock URL: \url{https://doi.org/10.1145/1399504.1360662}, \href {https://doi.org/10.1145/1399504.1360662} {\path{doi:10.1145/1399504.1360662}}.

\bibitem[CCK{\etalchar{*}}21]{Chen2021}
\textsc{Chen Z., Chen H.-Y., Kaufman D.~M., Skouras M., Vouga E.}:
\newblock Fine wrinkling on coarsely meshed thin shells.
\newblock \emph{ACM Trans. Graph. 40}, 5 (aug 2021).
\newblock URL: \url{https://doi.org/10.1145/3462758}, \href {https://doi.org/10.1145/3462758} {\path{doi:10.1145/3462758}}.

\bibitem[CCR{\etalchar{*}}20]{Casafranca2020}
\textsc{Casafranca J.~J., Cirio G., Rodríguez A., Miguel E., Otaduy M.~A.}:
\newblock {Mixing Yarns and Triangles in Cloth Simulation}.
\newblock \emph{Computer Graphics Forum} (2020).
\newblock \href {https://doi.org/10.1111/cgf.13915} {\path{doi:10.1111/cgf.13915}}.

\bibitem[CDHR08]{Chen2008Cholmod}
\textsc{Chen Y., Davis T.~A., Hager W.~W., Rajamanickam S.}:
\newblock Algorithm 887: Cholmod, supernodal sparse cholesky factorization and update/downdate.
\newblock URL: \url{https://doi.org/10.1145/1391989.1391995}, \href {https://doi.org/10.1145/1391989.1391995} {\path{doi:10.1145/1391989.1391995}}.

\bibitem[CKSV23]{Chen2023CWF}
\textsc{Chen Z., Kaufman D., Skouras M., Vouga E.}:
\newblock Complex wrinkle field evolution.
\newblock \emph{ACM Trans. Graph. 42}, 4 (July 2023).
\newblock URL: \url{https://doi.org/10.1145/3592397}, \href {https://doi.org/10.1145/3592397} {\path{doi:10.1145/3592397}}.

\bibitem[CKV19]{chen2019locking}
\textsc{Chen H.-Y., Kry P., Vouga E.}:
\newblock Locking-free simulation of isometric thin plates.
\newblock \emph{arXiv preprint arXiv:1911.05204} (2019).
\newblock URL: \url{https://arxiv.org/abs/1911.05204}, \href {https://doi.org/10.48550/arXiv.1911.05204} {\path{doi:10.48550/arXiv.1911.05204}}.

\bibitem[CLMMO14]{Cirio2014}
\textsc{Cirio G., Lopez-Moreno J., Miraut D., Otaduy M.~A.}:
\newblock Yarn-level simulation of woven cloth.
\newblock \emph{ACM Trans. Graph. 33}, 6 (Nov. 2014).
\newblock URL: \url{https://doi.org/10.1145/2661229.2661279}, \href {https://doi.org/10.1145/2661229.2661279} {\path{doi:10.1145/2661229.2661279}}.

\bibitem[CLMO15]{Cirio2015}
\textsc{Cirio G., Lopez-Moreno J., Otaduy M.~A.}:
\newblock Efficient simulation of knitted cloth using persistent contacts.
\newblock In \emph{Proceedings of the 14th ACM SIGGRAPH / Eurographics Symposium on Computer Animation} (New York, NY, USA, 2015), SCA '15, Association for Computing Machinery, p.~55–61.
\newblock URL: \url{https://doi.org/10.1145/2786784.2786801}, \href {https://doi.org/10.1145/2786784.2786801} {\path{doi:10.1145/2786784.2786801}}.

\bibitem[CLMO17]{Cirio2017}
\textsc{Cirio G., Lopez-Moreno J., Otaduy M.~A.}:
\newblock Yarn-level cloth simulation with sliding persistent contacts.
\newblock \emph{IEEE Transactions on Visualization and Computer Graphics 23}, 2 (Feb. 2017), 1152–1162.
\newblock URL: \url{https://doi.org/10.1109/TVCG.2016.2592908}, \href {https://doi.org/10.1109/TVCG.2016.2592908} {\path{doi:10.1109/TVCG.2016.2592908}}.

\bibitem[CT10]{Chen2010DevelopableCloth}
\textsc{Chen M., Tang K.}:
\newblock A fully geometric approach for developable cloth deformation simulation.
\newblock \emph{The Visual Computer 26}, 6-8 (2010), 853--863.
\newblock URL: \url{https://doi.org/10.1007/s00371-010-0467-5}, \href {https://doi.org/10.1007/s00371-010-0467-5} {\path{doi:10.1007/s00371-010-0467-5}}.

\bibitem[CXY{\etalchar{*}}23]{Chen2023:thick-shell}
\textsc{Chen Y., Xie T., Yuksel C., Kaufman D., Yang Y., Jiang C., Li M.}:
\newblock Multi-layer thick shells.
\newblock In \emph{ACM SIGGRAPH 2023 Conference Proceedings} (New York, NY, USA, 2023), SIGGRAPH '23, Association for Computing Machinery.
\newblock URL: \url{https://doi.org/10.1145/3588432.3591489}, \href {https://doi.org/10.1145/3588432.3591489} {\path{doi:10.1145/3588432.3591489}}.

\bibitem[DHSF00]{Doll2000}
\textsc{Doll S., Hauptmann R., Schweizerhof K., Freischl{\"a}ger C.}:
\newblock On volumetric locking of low-order solid and solid-shell elements for finite elastoviscoplastic deformations and selective reduced integration.
\newblock \emph{Engineering Computations 17}, 7 (2000), 874--902.
\newblock \href {https://doi.org/10.1108/02644400010355871} {\path{doi:10.1108/02644400010355871}}.

\bibitem[EB08]{English2008}
\textsc{English E., Bridson R.}:
\newblock Animating developable surfaces using nonconforming elements.
\newblock \emph{ACM Trans. Graph. 27}, 3 (Aug. 2008), 1–5.
\newblock URL: \url{https://doi.org/10.1145/1360612.1360665}, \href {https://doi.org/10.1145/1360612.1360665} {\path{doi:10.1145/1360612.1360665}}.

\bibitem[GHDS03]{Grinspun2003}
\textsc{Grinspun E., Hirani A.~N., Desbrun M., Schr\"{o}der P.}:
\newblock Discrete shells.
\newblock SCA '03, Eurographics Association, p.~62–67.

\bibitem[GJ{\etalchar{*}}10]{eigenweb}
\textsc{Guennebaud G., Jacob B., et~al.}:
\newblock Eigen v3.
\newblock \url{http://eigen.tuxfamily.org}, 2010.
\newblock Accessed: 2025-05-19.

\bibitem[GKS02]{Grinspun2002}
\textsc{Grinspun E., Krysl P., Schr\"{o}der P.}:
\newblock Charms: a simple framework for adaptive simulation.
\newblock \emph{ACM Trans. Graph. 21}, 3 (July 2002), 281–290.
\newblock URL: \url{https://doi.org/10.1145/566654.566578}, \href {https://doi.org/10.1145/566654.566578} {\path{doi:10.1145/566654.566578}}.

\bibitem[HCLK24]{Huang2024:GIPC}
\textsc{Huang K., Chitalu F.~M., Lin H., Komura T.}:
\newblock Gipc: Fast and stable gauss-newton optimization of ipc barrier energy.
\newblock \emph{ACM Trans. Graph. 43}, 2 (Mar. 2024).
\newblock URL: \url{https://doi.org/10.1145/3643028}, \href {https://doi.org/10.1145/3643028} {\path{doi:10.1145/3643028}}.

\bibitem[HS98]{Hauptmann1998}
\textsc{Hauptmann R., Schweizerhof K.}:
\newblock A systematic development of ‘solid-shell’ element formulations for linear and non-linear analyses employing only displacement degrees of freedom.
\newblock \emph{International Journal for Numerical Methods in Engineering 42}, 1 (1998), 49--69.
\newblock URL: \url{https://onlinelibrary.wiley.com/doi/abs/10.1002/(SICI)1097-0207(19980515)42:1<49::AID-NME349>3.0.CO;2-2}, \href {https://doi.org/10.1002/(SICI)1097-0207(19980515)42:1<49::AID-NME349>3.0.CO;2-2} {\path{doi:10.1002/(SICI)1097-0207(19980515)42:1<49::AID-NME349>3.0.CO;2-2}}.

\bibitem[HS02]{HARNAU2002805}
\textsc{Harnau M., Schweizerhof K.}:
\newblock About linear and quadratic “solid-shell” elements at large deformations.
\newblock \emph{Computers \& Structures 80}, 9 (2002), 805--817.
\newblock URL: \url{https://www.sciencedirect.com/science/article/pii/S0045794902000482}, \href {https://doi.org/https://doi.org/10.1016/S0045-7949(02)00048-2} {\path{doi:https://doi.org/10.1016/S0045-7949(02)00048-2}}.

\bibitem[JLGF17]{Jin2017}
\textsc{Jin N., Lu W., Geng Z., Fedkiw R.~P.}:
\newblock Inequality cloth.
\newblock In \emph{Proceedings of the ACM SIGGRAPH / Eurographics Symposium on Computer Animation} (New York, NY, USA, 2017), SCA '17, Association for Computing Machinery.
\newblock URL: \url{https://doi.org/10.1145/3099564.3099568}, \href {https://doi.org/10.1145/3099564.3099568} {\path{doi:10.1145/3099564.3099568}}.

\bibitem[KJM08]{Kaldor2008}
\textsc{Kaldor J.~M., James D.~L., Marschner S.}:
\newblock Simulating knitted cloth at the yarn level.
\newblock In \emph{ACM SIGGRAPH 2008 Papers} (New York, NY, USA, 2008), SIGGRAPH '08, Association for Computing Machinery.
\newblock URL: \url{https://doi.org/10.1145/1399504.1360664}, \href {https://doi.org/10.1145/1399504.1360664} {\path{doi:10.1145/1399504.1360664}}.

\bibitem[KJM10]{Kaldor2010}
\textsc{Kaldor J.~M., James D.~L., Marschner S.}:
\newblock Efficient yarn-based cloth with adaptive contact linearization.
\newblock \emph{ACM Trans. Graph. 29}, 4 (July 2010).
\newblock URL: \url{https://doi.org/10.1145/1778765.1778842}, \href {https://doi.org/10.1145/1778765.1778842} {\path{doi:10.1145/1778765.1778842}}.

\bibitem[KNO14]{Koh2014VDA}
\textsc{Koh W., Narain R., O'Brien J.~F.}:
\newblock View-dependent adaptive cloth simulation.
\newblock In \emph{Proceedings of the ACM SIGGRAPH/Eurographics Symposium on Computer Animation} (July 2014), pp.~1--8.
\newblock URL: \url{http://graphics.berkeley.edu/papers/Koh-VDA-2014-07/}.

\bibitem[LDN{\etalchar{*}}18]{Li2018:ARGUS}
\textsc{Li J., Daviet G., Narain R., Bertails-Descoubes F., Overby M., Brown G.~E., Boissieux L.}:
\newblock An implicit frictional contact solver for adaptive cloth simulation.
\newblock \emph{ACM Trans. Graph. 37}, 4 (July 2018).
\newblock URL: \url{https://doi.org/10.1145/3197517.3201308}, \href {https://doi.org/10.1145/3197517.3201308} {\path{doi:10.1145/3197517.3201308}}.

\bibitem[LFS{\etalchar{*}}20]{Li2020IPC}
\textsc{Li M., Ferguson Z., Schneider T., Langlois T., Zorin D., Panozzo D., Jiang C., Kaufman D.~M.}:
\newblock Incremental potential contact: Intersection- and inversion-free large deformation dynamics.
\newblock \emph{ACM Trans. Graph. (SIGGRAPH) 39}, 4 (2020).

\bibitem[LKJ21]{Li2021CIPC}
\textsc{Li M., Kaufman D.~M., Jiang C.}:
\newblock Codimensional incremental potential contact.
\newblock \emph{ACM Trans. Graph. (SIGGRAPH) 40}, 4 (2021).

\bibitem[LTT{\etalchar{*}}20]{Li2020Pcloth}
\textsc{Li C., Tang M., Tong R., Cai M., Zhao J., Manocha D.}:
\newblock P-cloth: interactive complex cloth simulation on multi-gpu systems using dynamic matrix assembly and pipelined implicit integrators.
\newblock \emph{ACM Trans. Graph. 39}, 6 (Nov. 2020).
\newblock URL: \url{https://doi.org/10.1145/3414685.3417763}, \href {https://doi.org/10.1145/3414685.3417763} {\path{doi:10.1145/3414685.3417763}}.

\bibitem[LWS{\etalchar{*}}18]{Leaf2018}
\textsc{Leaf J., Wu R., Schweickart E., James D.~L., Marschner S.}:
\newblock Interactive design of periodic yarn-level cloth patterns.
\newblock \emph{ACM Trans. Graph. 37}, 6 (Dec. 2018).
\newblock URL: \url{https://doi.org/10.1145/3272127.3275105}, \href {https://doi.org/10.1145/3272127.3275105} {\path{doi:10.1145/3272127.3275105}}.

\bibitem[MMCT24]{Montes2024q3t}
\textsc{Montes~Maestre J.~S., Coros S., Thomaszewski B.}:
\newblock Q3t prisms: A linear-quadratic solid shell element for elastoplastic surfaces.
\newblock In \emph{SIGGRAPH Asia 2024 Conference Papers} (New York, NY, USA, 2024), SA '24, Association for Computing Machinery.
\newblock URL: \url{https://doi.org/10.1145/3680528.3687697}, \href {https://doi.org/10.1145/3680528.3687697} {\path{doi:10.1145/3680528.3687697}}.

\bibitem[MMDH{\etalchar{*}}23]{Montes2023}
\textsc{Montes~Maestre J.~S., Du Y., Hinchet R., Coros S., Thomaszewski B.}:
\newblock Differentiable stripe patterns for inverse design of structured surfaces.
\newblock \emph{ACM Trans. Graph. 42}, 4 (July 2023).
\newblock URL: \url{https://doi.org/10.1145/3592114}, \href {https://doi.org/10.1145/3592114} {\path{doi:10.1145/3592114}}.

\bibitem[NSO12]{Narain2012ARCSim}
\textsc{Narain R., Samii A., O'Brien J.~F.}:
\newblock Adaptive anisotropic remeshing for cloth simulation.
\newblock \emph{ACM Trans. Graph. 31}, 6 (Nov. 2012).
\newblock URL: \url{https://doi.org/10.1145/2366145.2366171}, \href {https://doi.org/10.1145/2366145.2366171} {\path{doi:10.1145/2366145.2366171}}.

\bibitem[Qua12]{quaglino2012membrane}
\textsc{Quaglino A.}:
\newblock \emph{Membrane locking in discrete shell theories}.
\newblock Doctoral thesis, Georg-August-Universität Göttingen, 2012.
\newblock Accessed: 2025-05-19.
\newblock URL: \url{https://ediss.uni-goettingen.de/handle/11858/00-1735-0000-000D-F063-B}, \href {https://doi.org/10.53846/goediss-2533} {\path{doi:10.53846/goediss-2533}}.

\bibitem[Qua16]{Quaglino2016}
\textsc{Quaglino A.}:
\newblock A framework for creating low-order shell elements free of membrane locking.
\newblock \emph{International Journal for Numerical Methods in Engineering 108}, 1 (2016), 55--75.
\newblock URL: \url{https://onlinelibrary.wiley.com/doi/abs/10.1002/nme.5209}, \href {http://arxiv.org/abs/https://onlinelibrary.wiley.com/doi/pdf/10.1002/nme.5209} {\path{arXiv:https://onlinelibrary.wiley.com/doi/pdf/10.1002/nme.5209}}, \href {https://doi.org/https://doi.org/10.1002/nme.5209} {\path{doi:https://doi.org/10.1002/nme.5209}}.

\bibitem[SBRBO20]{Banderas2020}
\textsc{S\'{a}nchez-Banderas R.~M., Rodr\'{\i}guez A., Barreiro H., Otaduy M.~A.}:
\newblock Robust eulerian-on-lagrangian rods.
\newblock \emph{ACM Trans. Graph. 39}, 4 (Aug. 2020).
\newblock URL: \url{https://doi.org/10.1145/3386569.3392489}, \href {https://doi.org/10.1145/3386569.3392489} {\path{doi:10.1145/3386569.3392489}}.

\bibitem[SJLP11]{Sueda2011}
\textsc{Sueda S., Jones G.~L., Levin D. I.~W., Pai D.~K.}:
\newblock Large-scale dynamic simulation of highly constrained strands.
\newblock \emph{ACM Trans. Graph. 30}, 4 (July 2011).
\newblock URL: \url{https://doi.org/10.1145/2010324.1964934}, \href {https://doi.org/10.1145/2010324.1964934} {\path{doi:10.1145/2010324.1964934}}.

\bibitem[SNW20]{Sperl2020Homogenized}
\textsc{Sperl G., Narain R., Wojtan C.}:
\newblock Homogenized yarn-level cloth.
\newblock \emph{ACM Trans. Graph. 39}, 4 (Aug. 2020).
\newblock URL: \url{https://doi.org/10.1145/3386569.3392412}, \href {https://doi.org/10.1145/3386569.3392412} {\path{doi:10.1145/3386569.3392412}}.

\bibitem[SNW21]{Sperl2021}
\textsc{Sperl G., Narain R., Wojtan C.}:
\newblock Mechanics-aware deformation of yarn pattern geometry.
\newblock \emph{ACM Trans. Graph. 40}, 4 (July 2021).
\newblock URL: \url{https://doi.org/10.1145/3450626.3459816}, \href {https://doi.org/10.1145/3450626.3459816} {\path{doi:10.1145/3450626.3459816}}.

\bibitem[SSBL{\etalchar{*}}22]{Sperl2022}
\textsc{Sperl G., Sánchez-Banderas R.~M., Li M., Wojtan C., Otaduy M.~A.}:
\newblock Estimation of yarn-level simulation models for production fabrics.
\newblock \emph{ACM Transactions on Graphics (TOG) 41}, 4 (2022).

\bibitem[ST07]{Dimitri2007}
\textsc{Spillmann J., Teschner M.}:
\newblock {CORDE: Cosserat Rod Elements for the Dynamic Simulation of One-Dimensional Elastic Objects}.
\newblock In \emph{Eurographics/SIGGRAPH Symposium on Computer Animation} (2007), Metaxas D., Popovic J., (Eds.), The Eurographics Association.
\newblock \href {https://doi.org//10.2312/SCA/SCA07/063-072} {\path{doi:/10.2312/SCA/SCA07/063-072}}.

\bibitem[TF88]{Terzopoulos1988}
\textsc{Terzopoulos D., Fleischer K.}:
\newblock Modeling inelastic deformation: viscolelasticity, plasticity, fracture.
\newblock \emph{SIGGRAPH Comput. Graph. 22}, 4 (June 1988), 269–278.
\newblock URL: \url{https://doi.org/10.1145/378456.378522}, \href {https://doi.org/10.1145/378456.378522} {\path{doi:10.1145/378456.378522}}.

\bibitem[TG13]{Tamstorf2013}
\textsc{Tamstorf R., Grinspun E.}:
\newblock Discrete bending forces and their jacobians.
\newblock \emph{Graph. Models 75}, 6 (Nov. 2013), 362–370.
\newblock URL: \url{https://doi.org/10.1016/j.gmod.2013.07.001}, \href {https://doi.org/10.1016/j.gmod.2013.07.001} {\path{doi:10.1016/j.gmod.2013.07.001}}.

\bibitem[TTN{\etalchar{*}}13]{Tang2013}
\textsc{Tang M., Tong R., Narain R., Meng C., Manocha D.}:
\newblock A gpu-based streaming algorithm for high-resolution cloth simulation.
\newblock \emph{Computer Graphics Forum 32}, 7 (2013), 21--30.
\newblock URL: \url{https://onlinelibrary.wiley.com/doi/abs/10.1111/cgf.12208}, \href {http://arxiv.org/abs/https://onlinelibrary.wiley.com/doi/pdf/10.1111/cgf.12208} {\path{arXiv:https://onlinelibrary.wiley.com/doi/pdf/10.1111/cgf.12208}}, \href {https://doi.org/https://doi.org/10.1111/cgf.12208} {\path{doi:https://doi.org/10.1111/cgf.12208}}.

\bibitem[TwL{\etalchar{*}}18]{Tang2018}
\textsc{Tang M., wang t., Liu Z., Tong R., Manocha D.}:
\newblock I-cloth: incremental collision handling for gpu-based interactive cloth simulation.
\newblock \emph{ACM Trans. Graph. 37}, 6 (Dec. 2018).
\newblock URL: \url{https://doi.org/10.1145/3272127.3275005}, \href {https://doi.org/10.1145/3272127.3275005} {\path{doi:10.1145/3272127.3275005}}.

\bibitem[TWT{\etalchar{*}}16]{Tang2016}
\textsc{Tang M., Wang H., Tang L., Tong R., Manocha D.}:
\newblock {CAMA: Contact-Aware Matrix Assembly with Unified Collision Handling for GPU-based Cloth Simulation}.
\newblock \emph{Computer Graphics Forum} (2016).
\newblock \href {https://doi.org/10.1111/cgf.12851} {\path{doi:10.1111/cgf.12851}}.

\bibitem[Vou25]{vougalibshell}
\textsc{Vouga E.}:
\newblock libshell: A library for shell simulation and geometry processing.
\newblock \url{https://github.com/evouga/libshell}, 2025.
\newblock Accessed: 2025-05-19.

\bibitem[Wan21]{Wang:2021:GBS}
\textsc{Wang H.}:
\newblock Gpu-based simulation of cloth wrinkles at submillimeter levels.
\newblock \emph{ACM Trans. Graph. (SIGGRAPH) 40}, 4 (jul 2021).
\newblock URL: \url{https://doi.org/10.1145/3450626.3459787}, \href {https://doi.org/10.1145/3450626.3459787} {\path{doi:10.1145/3450626.3459787}}.

\bibitem[WOR10]{Wang2010}
\textsc{Wang H., O'Brien J., Ramamoorthi R.}:
\newblock Multi-resolution isotropic strain limiting.
\newblock In \emph{ACM SIGGRAPH Asia 2010 Papers} (New York, NY, USA, 2010), SIGGRAPH ASIA '10, Association for Computing Machinery.
\newblock URL: \url{https://doi.org/10.1145/1866158.1866182}, \href {https://doi.org/10.1145/1866158.1866182} {\path{doi:10.1145/1866158.1866182}}.

\bibitem[WWF{\etalchar{*}}18]{Wang2018}
\textsc{Wang Z., Wu L., Fratarcangeli M., Tang M., Wang H.}:
\newblock Parallel multigrid for nonlinear cloth simulation.
\newblock \emph{Computer Graphics Forum 37}, 7 (2018), 131--141.
\newblock URL: \url{https://onlinelibrary.wiley.com/doi/abs/10.1111/cgf.13554}, \href {http://arxiv.org/abs/https://onlinelibrary.wiley.com/doi/pdf/10.1111/cgf.13554} {\path{arXiv:https://onlinelibrary.wiley.com/doi/pdf/10.1111/cgf.13554}}, \href {https://doi.org/https://doi.org/10.1111/cgf.13554} {\path{doi:https://doi.org/10.1111/cgf.13554}}.

\bibitem[WWW22]{Wu:2022:GBM}
\textsc{Wu B., Wang Z., Wang H.}:
\newblock A gpu-based multilevel additive schwarz preconditioner for cloth and deformable body simulation.
\newblock \emph{ACM Trans. Graph. (SIGGRAPH) 41}, 4 (jul 2022).
\newblock URL: \url{https://doi.org/10.1145/3528223.3530085}, \href {https://doi.org/10.1145/3528223.3530085} {\path{doi:10.1145/3528223.3530085}}.

\bibitem[XTL19]{Xian2019}
\textsc{Xian Z., Tong X., Liu T.}:
\newblock A scalable galerkin multigrid method for real-time simulation of deformable objects.
\newblock URL: \url{https://doi.org/10.1145/3355089.3356486}, \href {https://doi.org/10.1145/3355089.3356486} {\path{doi:10.1145/3355089.3356486}}.

\bibitem[YSL{\etalchar{*}}24]{Yuan2024}
\textsc{Yuan C., Shi H., Lan L., Qiu Y., Yuksel C., Wang H., Jiang C., Wu K., Yang Y.}:
\newblock Volumetric homogenization for knitwear simulation.
\newblock \emph{ACM Transactions on Graphics (Proceedings of SIGGRAPH Asia 2024) 43}, 6 (12 2024), 207:1--207:19.
\newblock (*Joint First Authors).
\newblock URL: \url{https://doi.org/10.1145/3687911}, \href {https://doi.org/10.1145/3687911} {\path{doi:10.1145/3687911}}.

\bibitem[ZDF{\etalchar{*}}22]{Zhang2022}
\textsc{Zhang J.~E., Dumas J., Fei Y.~R., Jacobson A., James D.~L., Kaufman D.~M.}:
\newblock Progressive simulation for cloth quasistatics.
\newblock \emph{ACM Trans. Graph. 41}, 6 (2022).

\bibitem[ZDF{\etalchar{*}}23]{Zhang2023}
\textsc{Zhang J.~E., Dumas J., Fei Y.~R., Jacobson A., James D.~L., Kaufman D.~M.}:
\newblock Progressive shell quasistatics for unstructured meshes.
\newblock \emph{ACM Trans. Graph. 42}, 6 (2023).

\bibitem[ZJK24]{Zhang2024progressive}
\textsc{Zhang J.~E., James D.~L., Kaufman D.~M.}:
\newblock Progressive dynamics for cloth and shell animation.
\newblock \emph{ACM Transactions on Graphics (TOG) 43}, 4 (2024), 1--18.

\bibitem[ZLB{\etalchar{*}}24]{Zhang2024}
\textsc{Zhang J.~X., Lin G. W.-C., Bode L., Chen H.-Y., Stuyck T., Larionov E.}:
\newblock Estimating cloth elasticity parameters from homogenized yarn-level models.
\newblock In \emph{Proceedings of the 17th ACM SIGGRAPH Conference on Motion, Interaction, and Games} (New York, NY, USA, 2024), MIG '24, Association for Computing Machinery.
\newblock URL: \url{https://doi.org/10.1145/3677388.3696340}, \href {https://doi.org/10.1145/3677388.3696340} {\path{doi:10.1145/3677388.3696340}}.

\end{thebibliography}
